\documentstyle[11pt,aaspp4]{article}
\received{}
\accepted{}
\journalid{}{}
\articleid{}{}
\slugcomment{}

\lefthead{Veilleux et al.}
\righthead{Hidden BLRs in Ultraluminous Infrared Galaxies. II.}

\begin{document}

\title{New Results from a Near-Infrared Search for Hidden Broad-Line Regions 
in Ultraluminous Infrared Galaxies. }

\author{Sylvain Veilleux\altaffilmark{1,2}, D. B. Sanders\altaffilmark{3},
and D.-C. Kim\altaffilmark{4}}

\altaffiltext{1}{Department of Astronomy, University of Maryland,
College Park, MD 20742; E-mail: veilleux@astro.umd.edu}

\altaffiltext{2}{Visiting Astronomer at the United Kingdom Infrared Telescope, 
which is operated by the Royal Observatory Edingburgh on behalf of the U.K.
Science and Engineering Research Council}

\altaffiltext{3}{Institute for Astronomy, University of Hawaii, Honolulu, HI
96822; E-mail: sanders@.ifa.hawaii.edu}

\altaffiltext{4}{Infrared Processing and Analysis Center, 
California Institute of Technology, Pasadena, CA  91125;
E-mail: kim@ipac.caltech.edu}

\begin{abstract}

This paper reports the latest results from a near-infrared search for
hidden broad-line regions (BLRs: $\Delta V_{\rm FWHM} \ga$
2,000~km~s$^{-1}$) in ultraluminous infrared galaxies (ULIGs).  The
new sample contains thirty-nine ULIGs from the 1-Jy sample selected
for their lack of BLRs at optical wavelengths.  Broad Pa$\alpha$
emission is detected for the first time in two sources---F05189--2524
and F13305--1739
\footnote{Object names that begin with `F' are sources identified in
the {\em IRAS} Faint Source Catalog, Version 2\ (FSC: Moshir et al
1992)}. Broad Pa$\alpha$ emission may also be present in three other
sources---F13443+0802~SW, F14394+5332, and F16156+0146---but new data
are needed to make sure that H$_2$~$\lambda\lambda$1.8665, 1.8721 are
not contributing to this excess emission. The [Si~VI] feature, a
strong indicator of AGN activity, appears to be present in one
object---F13305--1739---and perhaps also in Mrk~273 and F13454--2956.
In addition, the presence of a hidden BLR is confirmed in the lower
luminosity source F11058--1131.

The results from this new study are combined with those from our
previously published survey (Veilleux, Sanders, \& Kim 1997b) to
produce a large database on 64 (non-Seyfert 1) ULIGs from the 1-Jy
sample.  All of the galaxies with strong evidence for a hidden BLR at
near-infrared wavelengths present an optical Seyfert~2 spectrum.
Overall, at least 50\% (and perhaps up to 70\%) of the optical
Seyfert~2 galaxies in the combined sample present either a BLR or
strong [Si~VI] emission.  In contrast, none of the 41 optically
classified LINERs and H~II galaxies in the sample shows any obvious
signs of an energetically important AGN. Galaxies with `warm' {\em
IRAS} colors ($f_{25}/f_{60} \ga 0.2$)\footnote{The quantities
$f_{25}$, $f_{60}$ are the {\em IRAS} flux densities in Jy at
25~\micron\ and 60~\micron\ , respectively.}  show a tendency to
harbor obscured BLRs in the near-infrared and to have large
Pa$\alpha$-to-infrared luminosity ratios.  These results support those
of our earlier survey and suggest that the screen of dust in most
`warm' Seyfert~2 galaxies is optically {\it thin} at 2~\micron.

When the results from this near-infrared survey are combined with
those from a recent optical spectroscopic study of the entire 1-Jy
sample of 118 ULIGs (Veilleux, Kim, \& Sanders 1999), we find that the
fraction of all ULIGs with optical or near-infrared signs of genuine
AGN activity (either a BLR or [Si~VI] emission) is at least 20
-- 25\%, but reaches 35 -- 50\% for objects with $L_{\rm ir} >
10^{12.3}\ L_\odot$.  Comparisons of the dereddened emission-line
luminosities of the optical or obscured BLRs detected in the ULIGs of
the 1-Jy sample with those of optical quasars indicate that the
obscured AGN/quasar in ULIGs is the main source of energy in at
least 15 -- 25\% of all ULIGs in the 1-Jy sample. This fraction is
closer to 30 -- 50\% among ULIGs with $L_{\rm ir} > 10^{12.3}\
L_\odot$. These results are compatible with those from recent
mid-infrared spectroscopic surveys carried out with {\it ISO}.

\end{abstract}


\section{Introduction}

In Veilleux, Sanders, \& Kim (1997b; hereafter VSK), a near-infrared
search for obscured broad-line regions (BLRs) and [Si~VI] emission
feature was carried out in a set of twenty-five ULIGs selected from
the {\em IRAS} 1-Jy survey of 118 ULIGs (Kim \& Sanders 1998).  Prior
to selecting their set of 25 objects, VSK excluded the $\sim$ 10\% of
the 118 galaxies that already were known optically to show direct
signs of quasar activity, i.e. optically classified as
Seyfert~1. Broad ($\Delta V_{\rm FWHM} \ga$ 2,000~km~s$^{-1}$)
infrared recombination lines were detected in 5 (possibly 8) of the 10
optically classified Seyfert 2 galaxies in the sample. The
high-ionization [Si~VI] feature ($\chi = 164$~eV), a clear signature
of an AGN, was detected in 3 or 4 of these Seyfert 2 galaxies.
Interestingly, all 6 of the `warm' ($f_{25}/f_{60} > 0.2$) optically
classified Seyfert 2 galaxies in their sample showed either obscured
BLRs or [Si~VI] emission at near-infrared wavelengths, as well as
large Pa$\alpha$-to-infrared luminosity ratios, strongly suggesting
that the screen of dust obscuring the cores of these ULIGs
is optically thin at 2~\micron.  In constrast, no obvious signs of an
obscured BLR or strong [Si~VI] emission were detected in any of the 15
optically classified LINERs and H~II galaxies in sample of VSK.  
The obscured BLRs detected in the Seyfert 2 galaxies were found to
have dereddened broad-line luminosities which are similar to those of
optically selected quasars of comparable bolometric luminosity. This
result was used by VSK to argue that most of the bolometric luminosity
in these broad-line objects is powered by the same mechanism as that
in optical quasars, namely mass accretion onto a supermassive black
hole.

The results of VSK rely on very small numbers of objects (e.g., there
were only 10 optically classified Seyfert 2 galaxies in their sample).
Their conclusions are therefore affected by large statistical
uncertainties.  For this reason and because of the success of this
initial survey, another set of 39 objects from the 1-Jy sample was
observed using the same techniques as VSK (cf.~Table 1).
F11058--1131, a luminous infrared galaxy (LIG) with
log(L$_{\rm IR}$/L$_{\odot}$) = 11.3, was also observed to test the
reported detection of broad polarized emission lines in this object
(Young et al. 1993).  The present paper describes the results of this
new study and combines them with the earlier data of VSK to produce a
sample of 64 ULIGs and 1 LIG which comprises 23 optically classified
Seyfert 2s, 25 LINERs, 16 H~II galaxies, and one object with
undetermined optical spectral type.

Section 2 of this paper describes the techniques used to acquire and
analyze the new data. The results of the new survey are presented in
Section 3 and combined with the results of VSK.  In Section 4, the
combined sample is used to test the conclusions of VSK.  The
conclusions from this analysis are summarized in Section 5 and
combined with our recent optical spectroscopic results.  This allows
us to make statistically meaningful statements about the fraction of
ULIGs which harbor a quasar, how this fraction varies with infrared
luminosity and optical spectral type, and whether the quasars detected
at optical and near-infrared wavelengths are energetically
significant.  A separate discussion of the emission-line properties of
each object in the new sample, with particular emphasis on the
presence or absence of broad components in the profiles of the
infrared hydrogen recombination lines, is presented in an Appendix.
We adopt H$_{\rm o}$ = 75 km s$^{-1}$ Mpc$^{-1}$ and q$_{\rm o}$ = 0
throughout this paper.

\section{Observations and Data Analysis}

To allow direct comparisons with the results of VSK, all of the new data
were acquired, reduced and analyzed using the same procedures as VSK
(see also Goodrich et al. 1994 and Veilleux, Goodrich, \& Hill 1997a).
All of these data were obtained with CGS4 on UKIRT (Mountain et
al. 1990).  A summary of the observations is presented in Table 2.
The 256 $\times$ 256 InSb array was used in conjunction with the 300 mm
focal length camera and the 75 l/mm grating. The slit width and pixel
scale were 1.$\arcsec$2. All of the spectra were obtained under
photometric conditions.  However, as described in VSK, the absolute
fluxes derived from these spectra are sensitive to slit losses,
seeing, centering errors, and guiding effects.  The line fluxes listed
in Table 3 have been corrected for slit losses (based on seeing
measurements taken throughout the night) but not for the other
effects. This source of errors is estimated to be less than about
50\%. A one-pixel (1.$\arcsec$2) extraction window was used to
transform the long-slit data into one-dimensional spectra. 

\section{Results}

The reduced spectra are presented in Figure 1 and discussed
individually in the Appendix.  Following the same format as in VSK,
Table 3 lists the fluxes, equivalent widths, and line widths of the
emission lines detected in these spectra. Also listed in this table
are the intensities of H$\alpha$ and H$\beta$, the line widths of
[O~III]~$\lambda$0.5007 and the color excess, $E(B-V)$, determined
from the optical emission-line flux ratios. Unless otherwise noted,
these measurements were taken from Kim et al. (1995), Kim, Veilleux,
\& Sanders (1998), Veilleux et al. (1995), or Veilleux, Kim, \&
Sanders (1999; hereafter VKS). The last two columns of Table 3 present
the color excesses determined from the infrared emission-line flux
ratios and based on the intrinsic flux ratios and extinction
coefficients of Veilleux et al. (1997a; their Table 2).

As found by VSK, the spectra of ULIGs are characterized by strong
Pa$\alpha$ emission and weaker lines from H$_2$.  Weak emission from
H$_2$ 1--0 S(5)~$\lambda$1.835, 1--0 S(2)~$\lambda$2.033, and 2--1
S(3)~$\lambda$2.073, and from Br$\gamma$~$\lambda$2.166,
Br$\delta$~$\lambda$1.945, Br$\epsilon$~$\lambda$1.817, and
He~I~$\lambda$2.058 is also visible in some galaxies (cf. Table
3). Figure 2 shows the distributions of the intensities of H$_2$ 1--0
S(3)~$\lambda$1.958 (possibly blended with [Si~VI]~$\lambda$1.962; see
below) and H$_2$ 1--0 S(1)~$\lambda$2.122 relative to narrow
Pa$\alpha$ as a function of optical spectral types. Ratios from the
combined set of objects in VSK and the present sample are
presented. This figure indicates that the H$_2$ emission in optical
H~II region-like galaxies is generally weaker than in optically
classified LINERs or Seyfert 2s.  The average H$_2$ 1--0
S(3)~$\lambda$1.958/Pa$\alpha$ ratios are 0.13, 0.23, and 0.32 for the
H~II galaxies (14 objects), LINERs (24) and Seyfert 2 galaxies (17),
respectively.  The average H$_2$ 1--0 S(1)~$\lambda$2.122/Pa$\alpha$
ratios are 0.11, 0.20, and 0.20 for the H~II galaxies (8 objects),
LINERs (15) and Seyfert 2 galaxies (10), respectively.
Kolmogorov-Smirnov (K-S) tests confirm that the difference between
the H~II galaxies and the other types of objects is significant.  The
H$_2$~$\lambda$1.958 and $\lambda$2.122 luminosities in the combined
sample of galaxies range from 5 $\times$ 10$^6$~$L_\odot$ to 3
$\times$ 10$^8$~$L_\odot$.  These luminosities translate into hot
H$_2$ masses of order 10,000--500,000~$M_\odot$ if the hot H$_2$
molecules are thermalized at T = 2,000 K (Scoville et al. 1982).

There are only two strong cases for broad ($\Delta V_{\rm FWHM} \ga$
2,000~km~s$^{-1}$) line emission in the new sample of ULIGs: F05189--2524
and F13305--1739. 
We also detect very broad ($\Delta V_{\rm FWHM} \sim$
7,200~km~s$^{-1}$) Pa$\alpha$ emission in the LIG F11058--1131,
therefore confirming the discovery of a hidden BLR in this object from
optical spectropolarimetry (Young et al. 1993).  In addition, many of
our objects present blue asymmetric wings at Pa$\alpha$ which can be
attributed to faint broad-line emission with $\Delta V_{\rm FWHM}
\approx$ 2,000--4,000~km s$^{-1}$.  However, in most cases (with the
possible exceptions of F04103--2838, F13443+0802~SW, F14394+5332, and
F16156+0146; see A4, A28, A33, and A35, respectively), this excess
emission on the blue-shifted side of Pa$\alpha$ is probably associated
with faint H$_2$ emission from the 7--5 O(3)~$\lambda$1.8721 and 6--4
O(5)~$\lambda$1.8665 transitions. As discussed in Black \&
van~Dishoeck (1987) and VSK, each of these lines is expected to be
about 1/3--1/4 the strength of the 1--0 S(3) transition if H$_2$ is
excited through fluorescence (Black \& van~Dishoeck 1987).  Combining
the current sample with that of VSK, we therefore find seven ULIGs
with obvious obscured BLRs.  All seven of these objects are optically
classified as Seyfert~2 galaxies, and all but one object (F23499+2423)
are `warm' galaxies with $f_{25}/f_{60} >$ 0.2 (the LIG F11058--1131
also is a Seyfert 2 galaxy with warm colors).

As in VSK, the resolution of our spectra is insufficient to deblend
the H$_2$ line at 1.958~\micron~from any possible
[Si~VI]~$\lambda$1.962 emission. Once again, we have to rely on the
fact that fluorescent excitation of molecular hydrogen at low density
predicts an H$_2$ 1--0 S(3)/1--0 S(1) flux ratio less than unity
(typically $\sim$~0.7; Black \& van Dishoeck 1987), while collisional
excitation predicts a ratio that ranges from 0.5 to 1.4 (Shull \&
Hollenbach 1978; Black \& van Dishoeck 1987; Sternberg \& Dalgarno
1989).  We conservatively assume that [Si~VI]~$\lambda$1.962 is
present in galaxies where we measure what appears to be H$_2$ 1--0
S(3)/1--0 S(1) flux ratios larger than 1.4.  As pointed out by VSK,
this lower limit is confirmed in active galaxies with measured
H$_2$~$\lambda$1.958 and $\lambda$2.122 line fluxes (e.g., Marconi et
al. 1994).  Only one object in the new sample clearly falls in this
category: the Seyfert 2 galaxy F13305--1739, one of the two ULIGs
in the new sample with broad Pa$\alpha$.  Two other Seyfert 2
galaxies, Mrk 273 and F13454--2956, may also present [Si~VI] emission,
but the evidence is less convincing. The H$_2$ 1--0 S(3) line
intensity of the LINERs F04103--2838 and F08572+3915NW appears
significantly larger than that of H$_2$ 1--0 S(1) but the H$_2$
$\lambda$1.958 feature is partially blended with Br$\delta$ in both
objects, making the intensity of this H$_2$ line uncertain.  If it is
assumed that the H$_2$ 1--0 S(3)/1--0 S(1) flux ratio in ULIGs is of
order unity---as generally seems to be the case in optically selected
Seyfert~2s (Marconi et al. 1994)---the [Si~VI]~$\lambda$1.962
luminosities of F13305--1739, Mrk 273, and F13454--2956 are: 15, 0.8,
and 2.0 $\times$ 10$^7$~$L_\odot$ (these luminosities have been
corrected for reddening using the color excesses derived from the
Pa$\alpha$/H$\alpha$ ratios listed in Table 3).  F13305--1739 is
therefore more luminous than F12072--0444, the most powerful [Si~VI]
emitter previously known (cf. VSK and Ward et al. 1991).

\section{Discussion}

Our new data confirm the tendency reported by VSK for the warmer
Seyfert 2 galaxies in the 1-Jy sample to reveal obscured BLRs at
near-infrared wavelengths.  All five objects in the combined sample of
ULIGs with $f_{25}/f_{60} > 0.33$ (Pks~13451+1232, F13305--1739,
Mrk~463E, F20460+1925, and F23060+0505) present obvious broad Paschen
emission lines (note that the LIG F11058--1131 would also fall in this
category). F05189--2524, another ULIG with unambiguous broad
Pa$\alpha$, is only marginally cooler than these objects
($f_{25}/f_{60}$ = 0.25). The only other object in our sample with
convincing broad Pa$\alpha$, F23499+2423, has $f_{25}/f_{60}$ = 0.12.
This BLR detection rate among Seyfert 2 ULIGs appears to be higher
than that found by Veilleux et al. (1997a) among optically-selected
Seyfert 2s. However, it is important to mention that the results on
the (low-redshift) optical Seyfert 2s relied on the detection of a
broad-line component to the weak Pa$\beta$ line rather than
Pa$\alpha$. Consequently, one cannot directly compare the results from
these two studies.

Our combined dataset allows us to make stronger statements on the
frequency of occurrence of nuclear activity among Seyfert 2
galaxies. Of the 22 ULIGs in the combined sample with optical
Seyfert~2 spectra, 10 of them show either obvious broad Pa$\alpha$ or
Pa$\beta$ emission, or strong evidence for [Si~VI]~$\lambda$1.962
emission. They are F05189--2524, F12072--0444, F13305--1739, Mrk~273,
Pks 1345+12, F13454--2956, Mrk~463E, F20460+1925, F23060+0505, and
F23499+2423 (recall that F11058--1131 is a LIG not a ULIG).  Six
additional Seyfert 2s from the combined sample may fall in this
category (F08559+1053, F13443+0802~SW, F14394+5332, F16156+0146,
F17179+5444, and F23233+2817), but data of higher spectral resolution
are needed to confirm the presence of [Si~VI] emission in these
objects or to determine the origin of the broad asymmetric wings to
their Pa$\alpha$ profiles. Near-infrared signs of nuclear activity are
therefore detected in 50 -- 70\% of the Seyfert 2 galaxies of the
combined sample.  These percentages are somewhat lower than those
found by VSK. The slight difference is probably due to the larger
fraction of `cool' ULIGs in the combined sample.  The combined
sample constitutes a more representative subset of the whole
(non-Seyfert 1) 1-Jy sample in terms of {\em IRAS} colors than the
original sample of VSK.  The results derived from the combined surveys
should therefore be used when generalizing to the entire 1-Jy sample

Our near-infrared results suggest that at least half of the optically
classified Seyfert 2 galaxies in the 1-Jy sample are genuine AGN. This
is a lower limit to the actual number since our near-infrared method
of AGN detection requires either the presence of a BLR or sufficient
ionizing radiation with energies above 164 eV to produce detectable
[Si~VI] $\lambda$1.962 emission. Some of the known optically selected
Seyfert~2 galaxies do not present this feature (cf. Marconi et
al. 1994).  Note that the BLR detection rate among Seyfert~2
galaxies appears to be lower than the 100\% (5 out of 5 or 6 out of 6
if the LIG F11058--1131 is also considered) detection rate among
objects with $f_{25}/f_{60}~>$ 1/3.  This stringent {\em IRAS}
25-to-60 $\mu$m color criterion is therefore a better indicator of BLR
activity in ULIGs than the Seyfert 2 characteristics at optical
wavelengths.

The high BLR detection rate among `warm' ULIGs suggests that the
screen of dust in these objects is often optically thin at 2~\micron.
Following VSK, we have plotted in Figure 3 the reddening-corrected
Pa$\alpha$ luminosities of our sample galaxies as a function of their
infrared luminosities.  The reddening correction was carried out using
the narrow-line extinctions listed in Table 3 and the lower limits on
the broad-line extinctions derived in the Appendix.  The solid line
represents the relation found by Goldader et al. (1997a, 1997b) among
lower luminosity infrared galaxies assuming a Pa$\alpha$/Br$\gamma$
ratio of 12 (case B recombination).  A significant fraction of the
data points fall below the solid line, therefore suggesting a deficit
in Pa$\alpha$ emission in many ULIGs. This result is different from
that of VSK, but is qualitatively similar to that of Goldader et
al. (1995), who suggested that this deficit is an indication that dust
obscuration is still important at 2~$\mu$m in some of these
objects. This apparent discrepancy with VSK is once again due to the
larger percentage of `cool' ULIGs in the combined sample.  Figure 4
illustrates this point.  The reddening-corrected Pa$\alpha$-to-IR
luminosity ratios of our sample galaxies are plotted as a function of
their {\em IRAS} $f_{25}/f_{60}$ ratio.  As first pointed out by VSK,
there is a clear tendency for the `warm' ULIGs to present larger
Pa$\alpha$-to-IR luminosity ratios than `cool' ULIGs.  Note that this
tendency is only visible when the Pa$\alpha$ fluxes include the
contribution from the broad-line component.

Our new results therefore confirm that the {\em IRAS} $f_{25}/f_{60}$
ratio is the primary factor determining the Pa$\alpha$-to-IR
luminosity ratio and the near-infrared detectability of obscured BLRs
in ULIGs. `Warm' ULIGs present optical and infrared properties which
are intermediate between those of `cool' ULIGs and optically selected
quasars. But are the putative `buried quasars' detected in these
objects powerful enough to provide the bulk of the energy output? This
question was first addressed by VSK using a sample of five objects
with obscured BLRs. VKS expanded this analysis by including ten
luminous and ultraluminous infrared galaxies with optically detected
BLRs (Seyfert 1s). Here, we further extend the analysis to include the
data from the three buried quasars/AGN detected in the new sample
(F05189--2524, F13305--1739, and the LIG F11058--131).

The results from this new analysis are presented in Figure 5, where
the dereddened emission-line luminosities of the optical and obscured
BLRs in infrared galaxies and in optically selected quasars are
plotted as a function of bolometric luminosity. A discussion of the
methods and assumptions which were used to create this figure is
presented in VKS and summarized in the caption to Figure 5. The
typical uncertainties on the data points of Figure 5 are of order
$\pm$ 30\%.  The solid line in the figure is the best log-linear fit
through the data of the optical quasars.  To first order, this line
therefore represents the relation found for pure broad-line
AGNs\footnote{Note, however, that even quasars may not be ``pure AGN''
since part of the far-infrared/submm emission may be produced by a
dusty starburst (Rowan-Robinson 1995). Here, we assume that all of the
far-infrared/submm emission is powered by the central AGN (Sanders et
al. 1989). The far-infrared/submm emission generally contributes about
20\% of the total bolometric luminosity of optical quasars.}.
Starburst-dominated ULIGs are expected to fall below this line. Figure
5 shows that F05189--2524 falls very close to the correlation derived
for optical quasars, while F11058--1131 and F13305--1739 lie less than
0.5 dex below the correlation. Our new results thus provide further
support to the idea first suggested by VSK and VKS that most ($\sim$
80\%) of the ULIGs with optical or near-infrared BLRs in the 1-Jy
sample are powered predominantly by the quasar rather than by a
powerful starburst. In other words, {\em the detection of an optical
or near-infrared BLR in a ULIG (about 20\% of the total 1-Jy sample)
appears to be an excellent sign that the AGN is the dominant energy
source in that ULIG.}

The good agreement in Figure 5 between optical quasars/Seyfert 1s and
Seyfert 2s with obscured BLRs is not expected within the standard
unification model of Seyfert galaxies. In this model, Seyfert 1s and
2s are fundamentally the same kind of objects and the viewing angle is
the only parameter of importance in determining their appearance (cf.,
e.g., Antonucci 1993). Consequently, Seyfert 1s and 2s are expected to
look the same when isotropic properties are compared.  One such
property is the far-infrared luminosity (since the far-infrared
emission is usually assumed to be optically thin thermal emission;
cf., e.g., Mulchaey et al. 1994 for a discussion of this issue).
Another isotropic property is the luminosity of the BLR {\em after}
correction for the extinction. The unification model therefore
predicts that quasars/Seyfert 1s and Seyfert 2s should follow the same
relation between the far-infrared luminosity and the
extinction-corrected broad-line H$\beta$ luminosity. The agreement
should break down when the bolometric luminosity is considered because
the near-infrared, big-blue bump, and X-ray emission that contributes
significantly to the bolometric luminosity of quasars/Seyfert 1s (but
not to that of Seyfert 2s) is supposedly radiated anisotropically.
This is not what we observe in Figure 5. Given that L$_{\rm IR}$
$\approx$ L$_{\rm BOL}$ in ULIGs and L$_{\rm IR}$ $<<$ L$_{\rm BOL}$
in quasars, Figure 5 implies that the BLRs of obscured Seyfert 1 ULIGs
are {\em underluminous} for a given far-infrared luminosity compared
to the quasars. Within the context of the unification model, one could
then interpret Figure 5 as evidence that the ULIGs produce a lot more
far-infrared emission than ``genuine'' AGN with the same broad-line
luminosity, thus that the AGN does {\em not} dominate their
energetics.

Alternatively, the standard unification model may not apply to ULIGs.  VKS
have already pointed out that the standard unification model has
difficulties explaining the larger percentage of Seyfert 1s relative
to Seyfert 2s among ULIGs of higher infrared luminosities.  It may be
that the obscuring mass distribution in ULIGs varies with
the luminosity of the energy source and with the optical spectral
type.  For instance, the trends with infrared luminosities can be
explained if the covering factor decreases with increasing infrared
luminosities (VKS). On the other hand, Figure 5 may indicate that
Seyfert 2 ULIGs have much larger dust covering factors (and therefore
reradiate a larger fraction of their bolometric luminosity in the far
infrared) than optical quasars and Seyfert 1 ULIGs. Perhaps the
obscuring material in Seyfert 2s has somehow had less time to settle
into a disk than the material in optical quasars/Seyfert 1s.

A recent comparison of our optical/near-infrared results with those
obtained with {\em ISO} (Lutz, Veilleux, \& Genzel 1999) lends support
to the scenario in which ULIGs with optical or obscured BLRs are
powered predominantly by an AGN. Lutz et al. (1999) conclude that
quasars constitute the dominant energy source in nearly all (10 out of
the 11) optically classified Seyfert ULIGs in common between the {\em
ISO} and optical samples.  Strong AGN activity, once triggered,
appears to quickly break the obscuring screen at least in certain
directions, thus becoming detectable over a wide wavelength range.

The situation in LINERs and H~II galaxies is more ambiguous.  These
objects represent about 70\% of the entire 1-Jy sample (VKS).  None of
the 41 optically classified LINERs and H~II galaxies in our
near-infrared sample show any obvious signs of an obscured BLR or
strong [Si~VI] emission\footnote{Broad Pa$\alpha$ may be present in
the LINER F04103--2838 (cf. A4), but spectra of higher spectral
resolution are needed to confirm this result. [Si~VI] $\lambda$1.962
emission may also be present in this object and the LINER F08572+3915
NW, but contamination by Br$\delta$ makes the intensity of this
feature uncertain. }. This apparent lack of an energetic AGN in these
objects is consistent with the recent {\it ISO} results (Genzel et
al. 1998; Lutz et 1999).  As discussed by VSK, two possible scenarios
can explain the near-infrared results on the LINERs and H~II galaxies:
(1) the cores of these ULIGs do not contain any AGN, (2) the optical
depth due to dust is large enough to hide the AGN even at 2
\micron. The broad range of {\em IRAS} colors and Pa$\alpha$-to-IR
luminosity ratios among LINERs and H~II galaxies (cf. Fig. 4) suggests
that these objects are affected by dust screens of various optical
depths.  Dust obscuration is therefore unlikely to be the {\em only}
reason why these objects present no obvious signs of AGN activity.

\section{Conclusions}

A sensitive near-infrared search for obscured broad-line regions was
carried out in a set of thirty-nine ULIGs selected from the {\em IRAS}
1-Jy survey of 118 ULIGs (Kim \& Sanders 1998).  The results of this
survey were combined with those obtained by VSK to produce a
near-infrared spectroscopic database on sixty-four ULIGs.  Prior to
selecting these objects, the $\sim$ 10\% of the 118 galaxies that
already were known optically to show direct signs of quasar activity,
i.e. optically classified as Seyfert~1, were excluded. The sixty-four
objects in the combined sample constitute a representative subset of
the whole (non-Seyfert 1) 1-Jy sample, in terms of redshift, infrared
luminosity, and {\em IRAS} colors. The results from the present survey
can therefore be generalized with a high degree of confidence to the
entire 1-Jy sample.  These results can be summarized as follows:

\begin{itemize}

\item[1.]  Broad infrared recombination lines are detected in seven
objects (all optically classified Seyfert 2s): F05198--2525,
F13305--1739, Pks~1345+12 (= F13451+1232), Mrk~463E (= F13536+1836),
F20460+1925, F23060+0505, and F23499+2423.  Excess broad emission is
also detected at the base of the Pa$\alpha$ profile in seven objects
(all but one are optically classified Seyfert 2s): F04103--2838 (the
only LINER), F08559+1053, F13443+0802~SW, F14394+5332, F16156+0146,
F17179+5444, and F23233+2817. But new data are required to determine
whether this emission is from H$_2$~$\lambda\lambda$1.8665,~1.8721
rather than from a genuine BLR. In addition, we confirm the
spectropolarimetric discovery of a hidden BLR in the optically
classified Seyfert 2 LIG F11058--1131.

\item[2.]  The [Si~VI] feature appears to be present in three
additional objects (all optically classified Seyfert 2 galaxies),
F12072--0444, Mrk~273 (= F13428+5608), and F13454--2956, and three or
quite possibly four of the objects mentioned above: F13305--1739,
Pks~1345+12, and F23233+2817, and perhaps F17179+5444.  If these
findings are confirmed by high-resolution spectroscopy, several of
these objects would be among the most luminous [Si~VI] emitters
presently known ($\ga$ 6 $\times$ 10$^7$~$L_\odot$).

\item[3.]  From results 1 and 2 above, we find that all nine `warm'
($f_{25}/f_{60} > 0.2$) optically classified Seyfert 2 galaxies in our
sample of ULIGs show either obscured BLRs or [Si~VI] emission at
near-infrared wavelengths. None of these objects presents deficient
Pa$\alpha$-to-infrared luminosity ratios.  These results support those
derived from the smaller sample of VSK, and suggest that the screen of
dust obscuring the cores of most `warm' Seyfert 2 ULIGs is optically
thin at 2~\micron.

\item[4.]  No obvious signs of an obscured BLR or strong [Si~VI]
emission are detected in any of the 41 optically classified LINERs and
H~II galaxies in our sample.  This result is consistent with recent
mid-infrared {\it ISO} observations of ULIGs. The LINERs and H~II
galaxies in our sample span a wide range of {\em IRAS} colors and
Pa$\alpha$-to-IR luminosity ratios. The apparent lack of AGN activity
in these objects is unlikely to be due solely to dust obscuration.

\end{itemize}

Recent results from an optical spectroscopic survey of the entire 1-Jy
sample of 118 ULIGs (VKS) indicate that about 10\% of these objects
are optically classified as Seyfert 1s, while about 20\% are optically
classified as Seyfert 2s. Above $L_{\rm ir} = 10^{12.3}\ L_\odot$, the
Seyfert 1s and 2s represent $\sim$ 26\% and 23\% of the ULIG
population, respectively.  Our near-infrared study of a representative
subset of 64 of these ULIGs indicates that {\em at least} 50 -- 70\%
(10 or possibly 16 out of 22) of the Seyfert 2s show signs of AGN
activity at rest wavelenths shortward of $\sim$ 2 $\mu$m (BLR or
[Si~VI] emission).  The optical and near-infrared data taken together,
therefore suggest that the total fraction of objects in the 1-Jy
sample with signs of a bonafide AGN is {\em at least} $\sim$ 20 --
25\%.  This fraction reaches 35 -- 50\% for objects with $L_{\rm ir} >
10^{12.3}\ L_\odot$.  These percentages are lower limits because the
near-infrared method often fails to detect AGN activity ([Si~VI]
emission) in known optically selected Seyfert 2 galaxies (Marconi et
al. 1994).

Comparisons of the dereddened emission-line luminosities of the
optical or obscured BLRs detected in the ULIGs of the 1-Jy sample with
those of optical quasars suggest that the AGN/quasar is the main
source of energy in $\sim$ 80\% of all ULIGs with optical or
near-infrared BLR. {\em At least} 15 -- 25\% of all ULIGs in the 1-Jy sample
are therefore powered primarily by the AGN/quasar rather than by a
starburst. This fraction is closer to 30 -- 50\% among ULIGs with
$L_{\rm ir} > 10^{12.3}\ L_\odot$. ULIGs with powerful AGN/quasar but
with no or highly obscured BLRs would increase these percentages.  
These results are consistent with those derived from {\em ISO} data.

\acknowledgments

S. V. and D. B. S. thank the organizers of the 1998 Ringberg meeting
where some of the issues in this paper were discussed, and the
referee, Tim Heckman, for his comments on the standard unification
model of Seyfert galaxies which helped improve this paper. This
research was supported in part by JPL contract no. 961566 to the
University of Hawaii (D. B. S.). S. V. gratefully acknowledges the
financial support of NASA through LTSA grant NAG 56547 and Hubble
fellowship HF-1039.01-92A awarded by the Space Telescope Science
Institute which is operated by the AURA, Inc. for NASA under contract
No. NAS5--26555. This research has made use of the NASA/IPAC
Extragalactic Database (NED) which is operated by the Jet Propulsion
Laboratory, California Institute of Technology, under contract with
NASA.

\clearpage

\centerline{APPENDIX}
\vskip 0.2in

\centerline{Notes on Individual Objects}
\vskip 0.1in

In this Appendix the emission-line properties of each object are
discussed, with particular emphasis on the presence or absence of
broad components in the profiles of the infrared hydrogen
recombination lines.  In the cases where no positive detection of
broad emission is reported, very conservative upper limits to the
broad-line fluxes are given. These broad-line fluxes were determined
by fitting the maximum broad Gaussian consistent with the data.

\vskip 0.2in
\centerline{A1. IRAS~F00188--0856}

An upper limit of $F$(Pa$\alpha_{\rm bl}$) $<$ 0.6 $\times$
 10$^{-14}$~ergs~s$^{-1}$~cm$^{-2}$, assuming $\Delta V_{\rm FWHM}
 \approx$ 7,000~km~s$^{-1}$, is derived in this optically classified
LINER.

\vskip 0.2in
\centerline{A2. IRAS~F03250+1606}

An upper limit of $F$(Pa$\alpha_{\rm bl}$) $<$ 1 $\times$
 10$^{-14}$~ergs~s$^{-1}$~cm$^{-2}$, assuming $\Delta V_{\rm FWHM}
 \approx$ 6,400~km~s$^{-1}$, is derived in this optically classified
 LINER.

\vskip 0.2in
\centerline{A3. IRAS~F04074--2801}

A wing is seen extending blueward from the base of Pa$\alpha$ in this
optically classified LINER.  A conservative fit to this wing emission
gives $F$(Pa$\alpha_{\rm bl}$) $\approx$ 0.5 $\times$
10$^{-14}$~ergs~s$^{-1}$~cm$^{-2}$ with $\Delta V_{\rm FWHM} \approx$
7,800~km~s$^{-1}$. However, contamination from
H$_2$~$\lambda\lambda$1.8665,~1.8721 may be responsible for this
feature.

\vskip 0.2in
\centerline{A4. IRAS~F04103--2838}

The optical spectrum of this object is ambiguous, spanning the
boundary between H~II galaxies and LINERs (VKS).  Interestingly, this
galaxy has the largest $f_{25}/f_{60}$ among all of the LINERs and
H~II galaxies of our combined sample ($f_{25}/f_{60}$ = 0.30). No
obvious broad Pa$\alpha$ emission is visible in the infrared spectrum
of this galaxy, but the Pa$\alpha$ profile presents a clear blue wing.
A conservative fit to this wing emission gives $F$(Pa$\alpha_{\rm
bl}$) $\approx$ 1.5 $\times$ 10$^{-14}$~ergs~s$^{-1}$~cm$^{-2}$ with
$\Delta V_{\rm FWHM} \approx$ 3,700~km~s$^{-1}$. Contamination from
H$_2$~$\lambda\lambda$1.8665,~1.8721 appears unlikely in this object
considering the weaknesses of the other H$_2$ lines. The feature near
1.96 $\mu$m is considerably stronger than H$_2$ $\lambda$2.122,
suggesting that [Si~VI] $\lambda$1.962 may be present in this
LINER. However, contamination from Br$\delta$ makes this statement
uncertain.

\vskip 0.2in
\centerline{A5. IRAS~F04313--1649}

This is the most distant source of our sample. There is no
high-resolution optical spectrum of this object that could be used to
determine the optical spectral type. This galaxy is relatively faint at
near-infrared wavelengths.  No obvious broad Pa$\alpha$ emission is
detected in this object.  An upper limit of $F$(Pa$\alpha_{\rm bl}$)
$<$ 0.2 $\times$ 10$^{-14}$~ergs~s$^{-1}$~cm$^{-2}$, assuming $\Delta
V_{\rm FWHM} \approx$~2,300~km~s$^{-1}$, is derived.

\vskip 0.2in
\centerline{A6. IRAS~F05024--1941}

The optical spectrum of this object is that of a Seyfert~2 galaxy
(VKS), but it has relatively `cool' {\em IRAS} 25-to-60 $\mu$m color
($f_{25}/f_{60}$ = 0.13). This object is faint at near-infrared
wavelengths. No obvious broad Pa$\alpha$ emission is detected in this
object within the uncertainties of our data.  An upper limit of
$F$(Pa$\alpha_{\rm bl}$) $<$ 0.3 $\times$
10$^{-14}$~ergs~s$^{-1}$~cm$^{-2}$, assuming $\Delta V_{\rm FWHM}
\approx$~2,500~km~s$^{-1}$, is derived.

\vskip 0.2in
\centerline{A7. IRAS~F05156--3024}

The optical spectrum of this object is that of a Seyfert~2 galaxy
(VKS), but it presents `cool' {\em IRAS} 25-to-60 $\mu$m color
($f_{25}/f_{60}$ = 0.089).  A weak wing is seen extending blueward
from the base of Pa$\alpha$ in this object.  A conservative fit to
this wing emission gives $F$(Pa$\alpha_{\rm bl}$) $\approx$ 0.5
$\times$ 10$^{-14}$~ergs~s$^{-1}$~cm$^{-2}$ with $\Delta V_{\rm FWHM}
\approx$ 2,400~km~s$^{-1}$. However, contamination from
H$_2$~$\lambda\lambda$1.8665,~1.8721 may be responsible for this
feature.

\vskip 0.2in
\centerline{A8. IRAS~F05189--2524}

This object is one of six ULIGs from the BGS sample of Sanders et
al. (1988a) included in the present study. This is the warmest object
of this set ($f_{25}/f_{60}$ = 0.25).  The optical spectrum of this
object is that of a Seyfert~2 galaxy (Veilleux et al. 1995).  The
K-band spectrum of this object clearly shows the presence of broad
($\Delta V_{\rm FWHM} \approx$ 2,600~km~s$^{-1}$) emission at
Pa$\alpha$. This spectrum also illustrates very well the difficulty in
detecting the broad component to Br$\gamma$. This explains the
negative results of Goldader et al. (1995) on this object. The J-band
spectrum confirms the presence of broad emission in the recombination
lines. Both Pa$\beta$ and He~I $\lambda$1.0832 present broad
profiles. Note, however, that the broad line parameters derived from
the J-band spectrum are considered less accurate than those derived
from Pa$\alpha$ because both Pa$\beta$ and He~I $\lambda$1.0832 are
weaker than Pa$\alpha$, and the intensities and profiles of both lines
are affected by strong atmospheric absorption features.  Based on the
broad Pa$\alpha$ flux listed in Table 3 and the assumption that a BLR
with one-third the intensity of the observed H$\alpha$ emission would
have been detected in the optical spectrum of VKS, a lower limit to
the color excess in the BLR of $E(B-V)_{\rm bl} >$~2.8~mag is derived.

\vskip 0.2in
\centerline{A9. IRAS~F08201+2801}

An upper limit of $F$(Pa$\alpha_{\rm bl}$) $<$ 0.2 $\times$
 10$^{-14}$~ergs~s$^{-1}$~cm$^{-2}$, assuming $\Delta V_{\rm FWHM}
 \approx$ 6,600~km~s$^{-1}$, is derived in this optically classified
 H~II galaxy.

\vskip 0.2in
\centerline{A10. IRAS~F08572+3915 NW}

This object is part of the original set of 10 ULIGs from the BGS
sample (Sanders et al. 1988a). The optical spectrum of this relatively
`warm' object ($f_{25}/f_{60}$ = 0.23) is ambiguous, spanning the
boundary between H~II galaxies and LINERs (Veilleux et al. 1995; VKS).
Our near-infrared spectrum does not reveal any obvious signs
of an obscured AGN (BLR or [Si~VI] feature) at near-infrared
wavelengths.  An upper limit of $F$(Pa$\alpha_{\rm bl}$) $<$ 1
$\times$ 10$^{-14}$~ergs~s$^{-1}$~cm$^{-2}$, assuming $\Delta V_{\rm
FWHM} \approx$ 6,700~km~s$^{-1}$, is derived.  A comparison of the
H$_2$~$\lambda$1.958 and $\lambda$2.122 fluxes suggests the presence
of [Si~VI]~$\lambda$1.962 emission in this galaxy, but this conclusion
is very uncertain because H$_2$~$\lambda$1.958 is clearly blended with
Br$\delta$.

\vskip 0.2in
\centerline{A11. IRAS~F10091+4704}

This powerful LINER/H~II galaxy presents very strong H$_2$ 1--0
S(3)~$\lambda$1.958 and 1--0 S(5) $\lambda$1.835 lines relative to
Pa$\alpha$. Weak [Fe~II] $\lambda$1.644 is also detected in this
object. The relatively poor signal-to-noise ratio of the spectrum near
Pa$\alpha$ (in part due to corrections for atmospheric absorption
features at those wavelengths) prevents us from making a strong
statement on the presence of a BLR in this object.  An upper limit of
$F$(Pa$\alpha_{\rm bl}$) $<$ 0.1 $\times$
10$^{-14}$~ergs~s$^{-1}$~cm$^{-2}$, assuming $\Delta V_{\rm FWHM}
\approx$ 5,800~km~s$^{-1}$, is derived.  We cannot make any 
statement on the existence of [Si~VI]~$\lambda$1.962 in this object
because the critical H$_2$ line at 2.122~\micron\ is not included
within the spectral range of our spectrum.

\vskip 0.2in
\centerline{A12. IRAS~F10190+1322}

This low-redshift H~II galaxy shows no obvious signs of obscured BLRs
or [Si~VI] $\lambda$1.962 emission. An upper limit of
$F$(Pa$\alpha_{\rm bl}$) $<$ 0.3 $\times$
10$^{-14}$~ergs~s$^{-1}$~cm$^{-2}$, assuming $\Delta V_{\rm FWHM}
\approx$ 7,300~km~s$^{-1}$, is derived.

\vskip 0.2in
\centerline{A13. IRAS~F10594+3818}

This object is optically classified as an H~II galaxy (VKS).  It shows
no obvious signs of obscured BLRs or [Si~VI] $\lambda$1.962
emission. An upper limit of $F$(Pa$\alpha_{\rm bl}$) $<$ 0.6 $\times$
10$^{-14}$~ergs~s$^{-1}$~cm$^{-2}$, assuming $\Delta V_{\rm FWHM}
\approx$ 7,300~km~s$^{-1}$, is derived.

\vskip 0.2in
\centerline{A14. IRAS~F11028+3130}

Our near-infrared spectrum of this LINER is essentially featureless.
An upper limit on the total Pa$\alpha$ flux of $F$(Pa$\alpha$)
$<$~0.04 $\times$ 10$^{-14}$~ergs~s$^{-1}$~cm$^{-2}$, assuming $\Delta
V_{\rm FWHM} \approx$ 500~km~s$^{-1}$, is derived. 

\vskip 0.2in
\centerline{A15. IRAS~F11095--0237}

The optical spectrum of this object is that of a LINER.  Our infrared
spectrum does not show any evidence for broad Pa$\alpha$ or [Si~VI]
$\lambda$1.962 emission. An upper limit of $F$(Pa$\alpha_{\rm bl}$)
$<$ 0.6 $\times$ 10$^{-14}$ ergs~s$^{-1}$~cm$^{-2}$, assuming $\Delta
V_{\rm FWHM} \approx$ 6,500~km~s$^{-1}$, is derived.

\vskip 0.2in
\centerline{A16. IRAS~F11180+1623}

There is no obvious signs of broad emission in the Pa$\alpha$ profile
of this LINER/H~II galaxy.  An upper limit of $F$(Pa$\alpha_{\rm bl}$)
$<$ 0.2 $\times$ 10$^{-14}$ ergs~s$^{-1}$~cm$^{-2}$, assuming $\Delta
V_{\rm FWHM} \approx$ 6,500~km~s$^{-1}$, is derived.  We cannot make
any strong statement on the existence of [Si~VI]~$\lambda$1.962 in this object
because the H$_2$ $\lambda$2.122 feature is close to the edge of the
spectrum and is strongly affected by atmospheric absorption features.

\vskip 0.2in
\centerline{A17. IRAS~F11223--1244}

There is no obvious signs for broad Pa$\alpha$ emission in this `cool'
($f_{25}/f_{60}$ = 0.11) Seyfert 2 galaxy. An upper limit of
$F$(Pa$\alpha_{\rm bl}$) $<$ 0.4 $\times$ 10$^{-14}$
ergs~s$^{-1}$~cm$^{-2}$, assuming $\Delta V_{\rm FWHM} \approx$
7,500~km~s$^{-1}$, is derived.  We cannot make any strong statement on
the existence of [Si~VI] $\lambda$1.962 in this galaxy because our
spectrum does not extend longward enough in wavelength to include the
critical H$_2$~$\lambda$2.122 line. An upper limit to the broad
Pa$\alpha$ flux of $F$(Pa$\alpha_{\rm bl}$) $<$ 0.3 $\times$
10$^{-14}$~erg~s$^{-1}$~cm$^{-2}$, assuming $\Delta V_{\rm FWHM}
\approx$ 5,000~km~s$^{-1}$, is derived.

\vskip 0.2in
\centerline{A18. IRAS~F11506+1331}

The optical spectrum of this object is ambiguous, spanning the
boundary between H~II galaxies and LINERs (VKS).  The
near-infrared spectrum is dominated by the strong Pa$\alpha$ feature.
The relatively weak H$_2$ lines are typical of those observed in
optically classified H~II galaxies (cf.~Fig.~1). There is no obvious
signs of AGN activity in this galaxy (BLR or [Si~VI] $\lambda$1.962
feature).  An upper limit of $F$(Pa$\alpha_{\rm bl}$) $<$ 1 $\times$
10$^{-14}$ ergs~s$^{-1}$~cm$^{-2}$, assuming $\Delta V_{\rm FWHM}
\approx$ 6,300~km~s$^{-1}$, is derived.

\vskip 0.2in
\centerline{A19. IRAS~F11582+3020}

No broad Pa$\alpha$ emission is detected in this LINER within the
uncertainties of our measurements.  An upper limit of
$F$(Pa$\alpha_{\rm bl}$) $<$ 0.2 $\times$ 10$^{-14}$
ergs~s$^{-1}$~cm$^{-2}$, assuming $\Delta V_{\rm FWHM} \approx$
3,600~km~s$^{-1}$, is derived.  The spectral range of our spectrum
prevents us from measuring the strength of H$_2$~$\lambda$2.122 and
comparing it with that of H$_2$~$\lambda$1.958.  The feature at
1.96~\micron\ is also too faint for detailed profile
analysis. Therefore, no statement can be made on the existence of
[Si~VI] $\lambda$1.962 in this galaxy.

\vskip 0.2in
\centerline{A20. IRAS~F12018+1941}

The optical spectral type of this galaxy (LINER) was derived using the
measurements of Armus et al. (1989). There is no evidence for [Si~VI]
$\lambda$1.962 or broad Pa$\alpha$ emission in this object. The continuum near
Pa$\alpha$ appears slightly convex, but this may be due to
contaminants redward of Pa$\alpha$. A very conservative upper limit to
the broad Pa$\alpha$ flux in this galaxy was derived by assuming that
the convexity in the continuum is due entirely to broad Pa$\alpha$
emission: $F$(Pa$\alpha_{\rm bl}$) $<$ 0.5 $\times$ 10$^{-14}$
ergs~s$^{-1}$~cm$^{-2}$, assuming $\Delta V_{\rm FWHM} \approx$
8,300~km~s$^{-1}$.

\vskip 0.2in
\centerline{A21. IRAS~F12032+1707}

This object shares many resemblances with F10091+4704 (cf.~A11). Both
of these galaxies are powerful (log[L$_{\rm IR}$/L$_\odot$] $>$ 12.5)
ULIGs with LINER characteristics at optical wavelengths and with
strong H$_2$ emission lines relative to Pa$\alpha$. The [Fe~II]
$\lambda$1.644 feature is also detected in both galaxies, and none of
these objects show any obvious signs for broad Pa$\alpha$. The quality
of our spectrum of F12032+1707 is slightly better than that of
F10091+4704, therefore allowing us to derive a strong upper limit to
the flux from broad Pa$\alpha$: $F$(Pa$\alpha_{\rm bl}$) $<$ 0.6
$\times$ 10$^{-14}$ ergs~s$^{-1}$~cm$^{-2}$, assuming $\Delta V_{\rm
FWHM} \approx$ 7,800~km~s$^{-1}$. No statement can be made about the
presence of [Si~VI] $\lambda$1.962 in this object because the H$_2$
$\lambda$2.122 is redshifted out of our spectral range.

\vskip 0.2in
\centerline{A22. IRAS~F12112+0305}

The optical spectrum of this BGS ULIG shows the signatures of a LINER.
The relatively low redshift of this object shifts the Pa$\alpha$ line
into a region of the spectrum which is affected by atmospheric
absorption features.  Nevertheless, the excellent correction for these
features (as evident from the flat and featureless continuum in this
region) allows us to derive a strong upper limit to the flux from
broad Pa$\alpha$: $F$(Pa$\alpha_{\rm bl}$) $<$ 0.7 $\times$
10$^{-14}$~ergs~s$^{-1}$~cm$^{-2}$ assuming $\Delta V_{\rm FWHM}
\approx$ 6,000~km~s$^{-1}$.  The H$_2$ lines in this object are all
very faint relative to Pa$\alpha$.

\vskip 0.2in
\centerline{A23. IRAS~F12447+3721}

The infrared spectrum of this H~II galaxy is similar to that of
F15206+3342, another H~II galaxy in the sample of VSK: strong narrow
emission from Pa$\alpha$ is visible superposed on a faint continuum;
the lines from H$_2$ are barely or not detected.  An upper limit to
the broad Pa$\alpha$ flux of $F$(Pa$\alpha_{\rm bl}$) $<$ 0.6 $\times$
10$^{-14}$~erg~s$^{-1}$~cm$^{-2}$, assuming $\Delta V_{\rm FWHM}
\approx$ 2,700~km~s$^{-1}$, is derived.

\vskip 0.2in
\centerline{A24. IRAS~F13106--0922}

The optical spectrum of this object is ambiguous, spanning the
boundary between H~II galaxies and LINERs (VKS).  No obvious signs of
[Si~VI] $\lambda$1.962 or broad Pa$\alpha$ emission are detected in
this object. An upper limit to the broad Pa$\alpha$ flux of
$F$(Pa$\alpha_{\rm bl}$) $<$ 0.3 $\times$
10$^{-14}$~erg~s$^{-1}$~cm$^{-2}$, assuming $\Delta V_{\rm FWHM}
\approx$ 5,000~km~s$^{-1}$, is derived.

\vskip 0.2in
\centerline{A25. IRAS~F13305--1739}

The infrared spectrum of this `warm' ($f_{25}/f_{60}$ = 0.34)
Seyfert~2 galaxy presents broad ($\Delta V_{\rm FWHM} \approx$
3,000~km~s$^{-1}$) Pa$\alpha$ emission but very little H$_2$ emission.
Based on the broad Pa$\alpha$ flux listed in Table 3 and the
assumption that a BLR with one-third the intensity of the observed
H$\alpha$ emission would have been detected in the optical spectrum of
VKS, a lower limit to the color excess in the BLR of $E(B-V)_{\rm bl}
>$~0.8~mag is derived.  Comparison of the H$_2$~$\lambda$1.958 and
$\lambda$2.122 fluxes also suggests the presence of
[Si~VI]~$\lambda$1.962 emission in this galaxy.

\vskip 0.2in
\centerline{A26. Mrk~273 = IRAS~F13428+5608}

This relatively `cool' ($f_{25}/f_{60}$ = 0.10) ULIG from the original
BGS sample (Sanders et al. 1988a) show Seyfert 2 characteristics at
optical wavelengths (Kim et al. 1998). Our J- and K-band spectra of
this object show no clear evidence for broad emission. The continuum
near Pa$\alpha$ appears slightly convex, but this may be due to our
correction for atmospheric absorption in this spectral region.  A very
conservative upper limit to the broad Pa$\alpha$ flux in this galaxy
was derived by assuming that the convexity in the continuum is due
entirely to broad Pa$\alpha$ emission: $F$(Pa$\alpha_{\rm bl}$) $<$
3.7 $\times$ 10$^{-14}$ ergs~s$^{-1}$~cm$^{-2}$, assuming $\Delta
V_{\rm FWHM} \approx$ 7,600~km~s$^{-1}$. An upper limit to the broad
Pa$\beta$ and He~I $\lambda$1.083 fluxes was also derived from the
J-band spectrum: $F$(He~I$_{\rm bl}$) $<$ 1.5 $\times$ 10$^{-14}$
ergs~s$^{-1}$~cm$^{-2}$, assuming $\Delta V_{\rm FWHM} \approx$
7,500~km~s$^{-1}$. Comparison of the H$_2$~$\lambda$1.958 and
$\lambda$2.122 fluxes suggests the presence of [Si~VI]~$\lambda$1.962
emission in this galaxy, but the evidence is marginal because the
H$_2$ $\lambda$1.958/H$_2$ $\lambda$2.122 flux ratio is only $\sim$
1.4 and contamination from Br$\delta$ makes the flux of H$_2$
$\lambda$1.958 uncertain.

\vskip 0.2in
\centerline{A27. IRAS~F13443+0802 NE}

The infrared spectrum of this H~II galaxy presents very weak H$_2$
emission lines and does not show any evidence for broad Pa$\alpha$
emission.  An upper limit to the broad Pa$\alpha$ flux of
$F$(Pa$\alpha_{\rm bl}$) $<$ 0.5 $\times$
10$^{-14}$~erg~s$^{-1}$~cm$^{-2}$, assuming $\Delta V_{\rm FWHM}
\approx$ 4,800~km~s$^{-1}$, is derived.

\vskip 0.2in
\centerline{A28. IRAS~F13443+0802 SW}

This is the galaxy in our sample with the lowest apparent Pa$\alpha$
flux.  Nevertheless, the S/N in the continuum is sufficient to put
strong constraints on any broad component to Pa$\alpha$. A weak blue
wing is detected at the base of Pa$\alpha$ in this Seyfert. A Gaussian
fit to this feature gives $F$(Pa$\alpha_{\rm bl}$) $\approx$ 0.2
$\times$ 10$^{-14}$ erg~s$^{-1}$~cm$^{-2}$ with a $\Delta V_{\rm FWHM}
\approx$ 2,400~km~s$^{-1}$.  Part of this feature may be attributed
to H$_2$~$\lambda\lambda$1.8665,~1.8721.  The line profile of
H$_2$~$\lambda$1.957 appears unusually broad ($\Delta V_{\rm FWHM}
\approx$ 1,250~km~s$^{-1}$), but it is uncertain.  The signal-to-noise
of our spectrum near H$_2$ $\lambda$2.122 is rather poor and prevents
us from making a strong statement on the relative intensity of H$_2$
$\lambda$1.958 and H$_2$ $\lambda$2.122.

\vskip 0.2in
\centerline{A29. IRAS~F13454--2956}

This Seyfert 2 galaxy has an unusually `cool' 25-to-60 $\mu$m IRAS flux
ratio of only 0.03.  There is no obvious signs of broad emission in
the profile of Pa$\alpha$. The excellent signal-to-noise ratio of our
near-infrared spectrum allows us to put strong constraints on any
broad component to Pa$\alpha$: $F$(Pa$\alpha_{\rm bl}$) $\approx$ 0.4
$\times$ 10$^{-14}$ erg~s$^{-1}$~cm$^{-2}$, assuming a $\Delta V_{\rm
FWHM} \approx$ 6,600~km~s$^{-1}$. Comparison of the H$_2$~$\lambda$1.958 and
$\lambda$2.122 fluxes suggests the presence of [Si~VI]~$\lambda$1.962
emission in this galaxy.

\vskip 0.2in
\centerline{A30. IRAS~F13509+0442}

The near-infrared spectrum of this H~II galaxy presents strong
Pa$\alpha$ but weak H$_2$. No obvious broad Pa$\alpha$ emission is
visible in our spectrum.  An upper limit of $F$(Pa$\alpha_{\rm bl}$)
$<$ 0.8 $\times$ 10$^{-14}$~ergs~s$^{-1}$~cm$^{-2}$, assuming $\Delta
V_{\rm FWHM} \approx$~4,000 km~s$^{-1}$, is derived.

\vskip 0.2in
\centerline{A31. IRAS~F14053--1958}

The only near-infrared emission line detected in this Seyfert~2 galaxy 
is narrow Pa$\alpha$. An upper limit of $F$(Pa$\alpha_{\rm bl}$)
$<$ 0.2 $\times$ 10$^{-14}$~ergs~s$^{-1}$~cm$^{-2}$, assuming $\Delta
V_{\rm FWHM} \approx$~2,300 km~s$^{-1}$, is derived.

\vskip 0.2in
\centerline{A32. IRAS~F14070+0525}

This object is the most powerful ULIG in our sample.  The infrared
spectrum of this Seyfert~2 galaxy is dominated by narrow
Pa$\alpha$. The [Fe~II] $\lambda$1.644 feature is also present at
considerably fainter flux levels.  An upper limit of
$F$(Pa$\alpha_{\rm bl}$) $<$ 0.2 $\times$
10$^{-14}$~ergs~s$^{-1}$~cm$^{-2}$, assuming $\Delta V_{\rm FWHM}
\approx$~3,800 km~s$^{-1}$, is derived.

\vskip 0.2in
\centerline{A33. IRAS~F14394+5332}

Several emission lines from H~I and H$_2$ are detected in the
near-infrared spectrum of this Seyfert 2 galaxy. A weak blue wing is
detected at the base of Pa$\alpha$. A Gaussian fit to this feature
gives $F$(Pa$\alpha_{\rm bl}$) $\approx$ 2 $\times$ 10$^{-14}$
erg~s$^{-1}$~cm$^{-2}$ with a $\Delta V_{\rm FWHM} \approx$
6,300~km~s$^{-1}$.  Part of this feature may be attributable to
H$_2$~$\lambda\lambda$1.8665,~1.8721. There is no evidence for strong
[Si~VI] emission in this object. 

\vskip 0.2in
\centerline{A34. IRAS~F15250+3609}

The optical spectrum of this BGS ULIG is ambiguous, straddling the
boundary between H~II galaxies and LINERs (VKS).  There is no obvious
signs for an obscured broad-line region or strong [Si~VI]
$\lambda$1.962 emission in this galaxy.  An upper limit of
$F$(Pa$\alpha_{\rm bl}$) $<$ 1.4 $\times$
10$^{-14}$~ergs~s$^{-1}$~cm$^{-2}$, assuming $\Delta V_{\rm FWHM}
\approx$~5,500 km~s$^{-1}$, is derived.

\vskip 0.2in
\centerline{A35. IRAS~F16156+0146}

A blue wing is detected at the base of Pa$\alpha$ in this optical
Seyfert 2. A Gaussian fit to this feature gives $F$(Pa$\alpha_{\rm
bl}$) $\approx$ 0.7 $\times$ 10$^{-14}$ erg~s$^{-1}$~cm$^{-2}$ with a
$\Delta V_{\rm FWHM} \approx$ 5,600~km~s$^{-1}$.  Part of this feature
may be attributable to H$_2$~$\lambda\lambda$1.8665,~1.8721.

\vskip 0.2in
\centerline{A36. IRAS~F16300+1558}

This optically classified LINER presents strong narrow Pa$\alpha$ and
weak H$_2$ $\lambda$1.958 and [Fe~II] $\lambda$1.644. An upper limit
of $F$(Pa$\alpha_{\rm bl}$) $<$ 0.3 $\times$
10$^{-14}$~ergs~s$^{-1}$~cm$^{-2}$, assuming $\Delta V_{\rm FWHM}
\approx$~3,500 km~s$^{-1}$, is derived.

\vskip 0.2in
\centerline{A37. IRAS~F17208--0014}

This BGS ULIG is optically classified as an H~II galaxy (Veilleux et
al. 1995).  There is no obvious broad Pa$\alpha$ emission in this
object within the uncertainties of the data. Errors in correcting for
the presence of atmospheric absorption features near Pa$\alpha$
increases the noise in this region.  A very conservative Gaussian fit
of the emission under narrow Pa$\alpha$ gives $F$(Pa$\alpha_{\rm bl}$)
$\approx$ 2 $\times$ 10$^{-14}$ erg~s$^{-1}$~cm$^{-2}$ with a $\Delta
V_{\rm FWHM} \approx$ 4,700~km~s$^{-1}$.

\vskip 0.2in
\centerline{A38. IRAS~F22491--1808}

This H~II galaxy is part of the original BGS sample of ULIGs (Sanders
et al.  1988).  The near-infrared spectrum of this object presents
only very weak H$_2$ emission and no evidence for strong [Si~VI] or
broad Pa$\alpha$ emission within the uncertainties of the data. Errors
in correcting for the presence of atmospheric absorption features near
Pa$\alpha$ increases the noise in this region.  A very conservative
Gaussian fit of the emission under narrow Pa$\alpha$ gives
$F$(Pa$\alpha_{\rm bl}$) $\approx$ 0.4 $\times$ 10$^{-14}$
erg~s$^{-1}$~cm$^{-2}$ with a $\Delta V_{\rm FWHM} \approx$
6,400~km~s$^{-1}$.

\vskip 0.2in
\centerline{A39. IRAS~F23365+3604}

The infrared spectrum of this optical LINER shows no evidence for an
obscured broad-line region or strong [Si~VI] $\lambda$1.962 emission.
An upper limit of $F$(Pa$\alpha_{\rm bl}$) $<$ 1 $\times$
10$^{-14}$~ergs~s$^{-1}$~cm$^{-2}$, assuming $\Delta V_{\rm FWHM}
\approx$~3,300 km~s$^{-1}$, is derived.

\vskip 0.2in
\centerline{A40. IRAS~F11058--1131 (LIG)}

Optical spectropolarimetry of this luminous infrared galaxy by Young
et al.  (1993) reveals a hidden BLR, seen in scattered, polarized
light.  Young et al. measure $\Delta V_{\rm FWHM} \approx$
7,600~km~s$^{-1}$ and $\Delta V_{\rm FWZI} \approx$ 16,800~km~s$^{-1}$
for polarized H$\alpha$. The infrared spectrum of this Seyfert 2
galaxy is equally stunning: broad ($\Delta V_{\rm FWHM} \approx$ 7,200
~km~s$^{-1}$) wings are clearly visible at Pa$\alpha$, extending to
$\sim$ $\pm$ 10,000~km~s$^{-1}$ at zero intensity and beautifully
confirming the results of Young et al. (1993). The ratio of
Pa$\alpha_{\rm bl}$ emission to scattered H$\alpha_{\rm bl}$ emission
is $\sim$ 8, nearly two orders of magnitude larger than the value
predicted from case B recombination.  If we make the rather safe
assumption that a BLR with one-third the intensity of the observed
H$\alpha$ emission would have been detected in the optical spectrum of
Osterbrock \& de Robertis (1985), we derive a lower limit to the color
excess towards the BLR based on Pa$\alpha_{\rm bl}$/H$\alpha_{\rm nl}$
of $E(B-V)_{\rm bl}$ $>$ 1.4.

\clearpage

% tables here
%%\documentstyle[aaspptwo]{article}
%\documentstyle[11pt,aaspp]{article}
%%\documentstyle[12pt,aasms]{article}
%%\received{Fall 1998}
%%\accepted{Winter 1998}
%%\journalid{}{}
%%\articleid{}{}
%%\slugcomment{}

%\begin{document}

\tiny

\begin{table*}
\tablenum{1}
\caption{Sample.} \label{tbl-1}

\begin{center}
\begin{tabular}{llrrrrrrrrl}
\tableline
\tableline
     &      &  \multicolumn{4}{c}{$f_\nu (\lambda)$~(Jy)} & {\rm FQ}$^a$ & $\frac{f_{25}}{f_{60}}$ & $\frac{f_{60}}{f_{100}}$ & $\frac{L_{ir}}{L_\odot}$ & Optical\\
IRAS       &  $z$  &  12\micron  &  25\micron  &  60\micron  &  100\micron  &   & log & log &  log  & Sp. Type\\
\tableline
\noalign{\vskip 2.5pt}  
F00188$-$0856      & 0.128 & 0.12 & 0.37 & 2.59 & 3.40 & 1332 & --0.85 & --0.12 & 12.33 & L\\
F03250+1606        & 0.129 & 0.10 & 0.15 & 1.38 & 1.77 & 1132 & --0.96 & --0.11 & 12.06 & L\\
F04074$-$2801      & 0.153 & 0.07 & 0.07 & 1.33 & 1.72 & 1a32 & --1.28 & --0.11 & 12.14 & L\\
F04103$-$2838      & 0.118 & 0.08 & 0.54 & 1.82 & 1.71 & 2332 & --0.53 &  +0.03 & 12.15 & L:\\
F04313$-$1649      & 0.268 & 0.07 & 0.07 & 1.01 & 1.10 & 1232 & --1.16 & --0.04 & 12.55 & ...\\
F05024$-$1941      & 0.192 & 0.15 & 0.14 & 1.06 & 1.34 & a232 & --0.88 & --0.10 & 12.43 & S2\\
F05156$-$3024      & 0.171 & 0.08 & 0.10 & 1.16 & 1.40 & 1232 & --1.06 & --0.08 & 12.20 & S2\\
F05189$-$2524      & 0.042 & 0.73 & 3.44 &13.67 &11.36 & 3332 & --0.60 &  +0.08 & 12.07 & S2\\
F08201+2801        & 0.168 & 0.09 & 0.15 & 1.17 & 1.43 & 1a32 & --0.89 & --0.09 & 12.23 & H\\
F08572+3915 NW     & 0.058 & 0.32 & 1.70 & 7.43 & 4.59 & 3332 & --0.64 &  +0.21 & 12.11 & L:\\
F10091+4704        & 0.246 & 0.06 & 0.08 & 1.18 & 1.55 & 1132 & --1.17 & --0.12 & 12.67 & L:\\
F10190+1322        & 0.077 & 0.07 & 0.38 & 3.33 & 5.57 & 1232 & --0.94 & --0.22 & 12.00 & H\\
F10594+3818        & 0.158 & 0.09 & 0.15 & 1.29 & 1.89 & 1132 & --0.93 & --0.17 & 12.24 & H\\
F11028+3130        & 0.199 & 0.09 & 0.09 & 1.02 & 1.44 & 1a32 & --1.05 & --0.15 & 12.32 & L\\
F11095$-$0237      & 0.106 & 0.06 & 0.42 & 3.25 & 2.53 & a332 & --0.89 &  +0.11 & 12.20 & L\\
F11180+1623        & 0.166 & 0.08 & 0.19 & 1.19 & 1.60 & 1132 & --0.80 & --0.13 & 12.24 & L:\\
F11223$-$1244      & 0.199 & 0.07 & 0.16 & 1.52 & 2.26 & 1132 & --0.98 & --0.17 & 12.59 & S2\\
F11506+1331        & 0.127 & 0.10 & 0.19 & 2.58 & 3.32 & 1a32 & --1.13 & --0.11 & 12.28 & H:\\
F11582+3020        & 0.223 & 0.10 & 0.15 & 1.13 & 1.49 & 1132 & --0.88 & --0.12 & 12.56 & L\\
F12018+1941        & 0.169 & 0.11 & 0.37 & 1.76 & 1.78 & 1232 & --0.68 &  +0.00 & 12.44 & L$^b$\\
F12032+1707        & 0.217 & 0.14 & 0.25 & 1.36 & 1.54 & 1a32 & --0.74 & --0.05 & 12.57 & L\\
F12112+0305        & 0.073 & 0.12 & 0.51 & 8.50 & 9.98 & a332 & --1.22 & --0.07 & 12.28 & L\\
F12447+3721        & 0.158 & 0.12 & 0.10 & 1.04 & 0.84 & 1a32 & --0.94 &  +0.09 & 12.06 & H\\
F13106$-$0922      & 0.174 & 0.12 & 0.06 & 1.24 & 1.89 & 1122 & --1.32 & --0.18 & 12.32 & L:\\
F13305$-$1739      & 0.148 & 0.09 & 0.39 & 1.16 & 1.04 & 1232 & --0.47 &  +0.05 & 12.21 & S2\\
F13428+5608$^b$& 0.037 & 0.24 & 2.28 &21.74 &21.38 & 3332 & --0.98 &  +0.01 & 12.10 & S2\\
F13443+0802 NE/SW  & 0.135 & 0.12 & 0.11 & 1.50 & 1.99 & 1132 & --1.07 & --0.12 & 12.15 & H/S2\\
F13454$-$2956      & 0.129 & 0.06 & 0.07 & 2.16 & 3.38 & 1a32 & --1.49 & --0.19 & 12.21 & S2\\
F13509+0442        & 0.136 & 0.10 & 0.23 & 1.56 & 2.53 & 2132 & --0.83 & --0.21 & 12.27 & H\\
F14053$-$1958      & 0.161 & 0.07 & 0.14 & 1.02 & 1.12 & 1132 & --0.86 & --0.04 & 12.12 & S2\\
F14070+0525        & 0.265 & 0.07 & 0.19 & 1.45 & 1.82 & 1a32 & --0.88 & --0.10 & 12.76 & S2\\
F14394+5332        & 0.105 & 0.03 & 0.35 & 1.95 & 2.39 & a332 & --0.75 & --0.09 & 12.04 & S2\\
F15250+3609$^d$    & 0.055 & 0.20 & 1.32 & 7.29 & 5.91 & 1332 & --0.74 &  +0.09 & 11.97 & L:\\
F16156+0146        & 0.132 & 0.10 & 0.28 & 1.13 & 1.00 & 1332 & --0.61 &  +0.05 & 12.04 & S2\\
F16300+1558        & 0.242 & 0.07 & 0.07 & 1.48 & 1.99 & 1a32 & --1.33 & --0.13 & 12.63 & L\\
F17208$-$0014$^d$  & 0.043 & 0.20 & 1.66 &31.14 &34.90 & 2332 & --1.27 & --0.05 & 12.39 & H\\
F22491$-$1808      & 0.076 & 0.05 & 0.55 & 5.44 & 4.45 & a232 & --1.00 &  +0.09 & 12.09 & H\\
F23365+3604$^d$    & 0.064 & 0.10 & 0.81 & 7.09 & 8.36 & 1332 & --0.94 & --0.07 & 12.10 & L\\
\noalign{\vskip 4.0pt}  
F11058$-$1131$^d$ (LIG)& 0.055$^e$ & 0.15 & 0.32 & 0.77 & 0.79 & 1232 & --0.38 & --0.01 & 11.27 & S2$^e$\\
\noalign{\vskip 2.5pt}  
\tableline
\end{tabular}
\end{center}

% Text for table footnotes must follow the tabular environment but must
% be inside the table environment.  Note that it is OK to put \ref's
% in \tablenotetext's.
\tablenotetext{^a}{Flux quality in each of the four {\em IRAS}
bands. Numbers 1-3 are the quality flags adopted for the {\em IRAS}
catalogs (1 = upper limit, 2 = moderate quality, 3 = high
quality). The letter `a' means that the flux density was estimated by
Kim \& Sanders (1998) after coadding all of the {\em IRAS} data
available for the source using the ADDSCAN/SCANPI procedure (Helou et
al. 1988); these `low quality' measurements are typically 2-3 $\sigma$
detections.  }
\tablenotetext{^b}{Based on the line ratios measured by Armus,
Heckman, \& Miley (1989). Only a lower limit is available on the
[O~III] $\lambda$0.5007/H$\beta$ ratio. It is therefore possible that
this object is a Seyfert 2 galaxy instead of a LINER. }
\tablenotetext{^c}{Other name: F13428+5608 = Mrk~273.}
\tablenotetext{^d}{Infrared properties from the {\em IRAS} Faint
Source Catalog, Version 2 (Moshir et al. 1992).}
\tablenotetext{^e}{Redshift and optical spectral type from Osterbrock
\& de Robertis (1985).}

\tablecomments{Unless otherwise noted, infrared properties and optical
spectral types are from Kim \& Sanders (1998), Kim, Veilleux, \&
Sanders (1998), or Veilleux et al (1995, 1999).}

\end{table*}

\clearpage

%\end{document}

%%\documentstyle[aaspptwo]{article}
%\documentstyle[11pt,aaspp]{article}
%%\documentstyle[12pt,aasms]{article}
%%\received{Fall 1998}
%%\accepted{Winter 1998}
%%\journalid{}{}
%%\articleid{}{}
%%\slugcomment{}

%\begin{document}

\tiny

\begin{table*}
\tablenum{2}
\caption{Journal of Observations.} \label{tbl-2}

\begin{center}
\begin{tabular}{lrrc}
\tableline
\tableline
&  & Exp & Res\\
IRAS & Date & (min) & ($\frac{\lambda}{\Delta\lambda}$)\\
(1) & (2) & (3) & (4) \\
\tableline
\noalign{\vskip 2.5pt}  
F00188$-$0856  & 1997 Dec 27   &  64&  430\\
F03250+1606    & 1997 Dec 28   &  32&  430\\
F04074$-$2801  & 1997 Dec 28   &  64&  430\\
F04103$-$2838  & 1997 Dec 28   &  16&  430\\
F04313$-$1649  & 1998 Mar 05   &  80&  400\\
F05024$-$1941  & 1997 Dec 27   &  64&  430\\
F05156$-$3024  & 1997 Dec 27   &  32&  430\\
F05189$-$2524  & 1997 Dec 27   &  16&  430\\
               & 1997 Dec 28   &  16&  430\\
               & 1997 Dec 28   &  24&  500\\
F08201+2801    & 1997 Dec 28   &  48&  430\\
F08572+3915 NW & 1997 Dec 27   &  40&  430\\
F10091+4704    & 1997 Dec 27   &  80&  430\\
F10190+1322    & 1998 Mar 04   &  64&  400\\
F10594+3818    & 1998 Mar 04   &  32&  400\\
F11028+3130    & 1998 Mar 05   &  64&  400\\
F11095$-$0237  & 1998 Mar 04   &  48&  400\\
F11180+1623    & 1998 Mar 04   &  48&  400\\
F11223$-$1244  & 1997 Dec 27   &  64&  430\\
F11506+1331    & 1998 Mar 04   &  16&  400\\
F11582+3020    & 1997 Dec 28   &  64&  430\\
F12018+1941    & 1998 Mar 04   &  32&  400\\
F12032+1707    & 1998 Mar 05   &  32&  400\\
F12112+0305    & 1997 Dec 27   &  32&  430\\
               & 1997 Dec 28   &  32&  430\\
F12447+3721    & 1998 Mar 07   &  24&  400\\
F13106$-$0922  & 1998 Mar 05   &  64&  400\\
F13305$-$1739  & 1998 Mar 04   &  32&  400\\
F13428+5608    & 1998 Mar 04   &  16&  400\\
               & 1998 Mar 07   &  16&  460\\
F13443+0802 NE & 1998 Mar 06   &  24&  400\\
F13443+0802 SW & 1998 Mar 06   &  48&  400\\
F13454$-$2956  & 1998 Mar 06   &  48&  400\\
F13509+0442    & 1998 Mar 04   &  24&  400\\
F14053$-$1958  & 1998 Mar 06   &  40&  400\\
               & 1998 Mar 07   &  32&  400\\
F14070+0525    & 1998 Mar 04   &  32&  400\\
               & 1998 Mar 06   &  48&  400\\
F14394+5332    & 1998 Mar 05   &  16&  400\\
F15250+3609    & 1998 Mar 05   &  16&  400\\
F16156+0146    & 1998 Mar 05   &  48&  400\\
F16300+1558    & 1998 Mar 07   &  40&  400\\
F17208$-$0014  & 1998 Mar 05   &  16&  400\\
F22491$-$1808  & 1997 Dec 28   &  48&  430\\
F23365+3604    & 1997 Dec 28   &  32&  430\\
\noalign{\vskip 4.0pt}  
F11058$-$1131 (LIG) & 1998 Mar 05   &  16&  400\\
\noalign{\vskip 2.5pt}  
\tableline
\end{tabular}
\end{center}

% Text for table footnotes must follow the tabular environment but must
% be inside the table environment.  Note that it is OK to put \ref's
% in \tablenotetext's.

\end{table*}

\clearpage

%\end{document}

%%\documentstyle[aaspptwo]{article}
%\documentstyle[11pt,aaspp]{article}
%%\documentstyle[12pt,aasms]{article}
%%\received{Fall 1998}
%%\accepted{Winter 1998}
%%\journalid{}{}
%%\articleid{}{}
%%\slugcomment{}

%\begin{document}

\tiny

\begin{table*}
\tablenum{3}
\caption{Optical$^a$ and Near-Infrared Line Measurements.} \label{tbl-3}

\begin{center}
\begin{tabular}{llccccccccccccccc}
\tableline
\tableline
IRAS &  & H$\beta_{\rm nl}$ & H$\alpha_{\rm nl}$ & [Fe~II] & H$_2$ & Pa$\alpha_{\rm nl}$ & Pa$\alpha_{\rm bl}$ & 
Br$\delta$ & H$_2$ 1.958 & H$_2$ &He I &H$_2$ & Br$\gamma$ & \multicolumn{3}{c}{E(B--V)$_{\rm nl}$}   \\
\cline{15-17}\\
 & & & & 1.644 & 1.835 & 1.875&  1.875& 1.945 & + [Si~VI] 1.962 & 2.033 & 2.058 & 2.121 &  & H$\alpha$/H$\beta$  & 
 Pa$\alpha$/H$\alpha$ & Br$\gamma$/H$\alpha$\\
(1) & (2) & (3) & (4) & (5) & (6) & (7) & (8) & (9) & (10) & (11) & (12) & (13) & (14) & (15) & (16) & (17)\\
\tableline
\noalign{\vskip 7.5pt}  
%name                   hb    ha  1.644  1.835 pa_nl pa_bl  brd  1.958 2.033 2.058 2.121  brg  ha/hb pa/ha brg/ha  
F00188$-$0856& $F$&  0.0245& 0.35&   ...& 0.109&0.907&  ...&  ...&0.204&  ...&  ...& 0.4:& 0.2:& 1.56& 1.72& 2.15: \\ 
             & $EW$&       &     &   ...&   6.0&   49&  ...&  ...& 11.3&  ...&  ...&  21:&  12:& & & \\            
             & $W$&        &     &   ...&  689:&  831&  ...&  ...&  875&  ...&  ...&1200::&1320::& & & \\            
             & $W_c$&      &  ...&   ...&    0:&  448&  ...&  ...&  526&  ...&  ...& 975::&1120::& & & \\            
\noalign{\vskip 7.5pt}  
F03250+1606  & $F$&  0.0220& 0.44&   ...&   ...& 3.00&  ...&  ...&0.311&  ...&0.11:  &0.20:&0.23:& 1.93& 2.24& 2.11: \\ 
	     & $EW$&       &     &   ...&   ...&  126&  ...&  ...& 15.2&  ...& 5.9:  &10.2:&12.4:& & & \\            
	     & $W$&        &     &   ...&   ...&  813&  ...&  ...&  884:&  ...& 570::& 611::& 804::& & & \\            
	     & $W_c$&      & 320:&   ...&   ...&  414&  ...&  ...&  539:&  ...&   0::&   0::& 395::& & \\            
\noalign{\vskip 7.5pt}  
F04074$-$2801& $F$&  0.0077& 0.07&   ...& 0.138&0.459&  ...&  ...&0.192&  ...&  ...&0.275&  ...& 1.12& 2.22&... \\ 
	     & $EW$&       &     &   ...&  13.7& 47.5&  ...&  ...& 22.5&  ...&  ...& 38.4&  ...& & & \\            
	     & $W$&        &     &   ...&   961:&  803&  ...&  ...&  788&  ...&  ...&  882&  ...& & & \\            
	     & $W_c$&      &  ...&   ...&   658:&  392&  ...&  ...&  362&  ...&  ...&  536&  ...& & & \\            
\noalign{\vskip 7.5pt}  
F04103$-$2838& $F$&   0.128& 1.42&   ...&   ...& 3.37&  ...&0.17:&0.393&  ...&  ...&0.14:&0.441& 1.25& 1.67& 1.84 \\ 
	     & $EW$&       &     &   ...&   ...&  166&  ...& 8.1:& 19.5&  ...&  ...& 7.1:& 22.8& & & \\            
	     & $W$&        &     &   ...&   ...&  827&  ...& 754::& 1092:&  ...&  ...& 447::&  717:& & & \\            
	     & $W_c$&      &  390&   ...&   ...&  440&  ...& 279::&  839:&  ...&  ...&   0::&  154:& & & \\            
\noalign{\vskip 7.5pt}  
F04313$-$1649& $F$&     ...&  ...&   ...&   ...&0.219&  ...&  ...&  ...&  ...&  ...&  ...&  ...& ...& ...&... \\ 
	     & $EW$&       &     &   ...&   ...& 53.4&  ...&  ...&  ...&  ...&  ...&  ...&  ...& & & \\            
	     & $W$&        &     &   ...&   ...&  642&  ...&  ...&  ...&  ...&  ...&  ...&  ...& & & \\            
	     & $W_c$&      &  ...&   ...&   ...&    0&  ...&  ...&  ...&  ...&  ...&  ...&  ...& & & \\            
\noalign{\vskip 7.5pt}  
F05024$-$1941& $F$& 0.0084:&0.21:&0.08::& 0.067:& 0.414& ...&0.074::& 0.102&  ...&  ...&  ...&  ...& 2.13& 1.57:&... \\ 
	     & $EW$&       &     &   6::&  6.0:&    38& ...&   7.5::&  10.7&  ...&  ...&  ...&  ...& & & \\            
	     & $W$&        &     &   ...&  708::&  862& ...&  ...&    640:&  ...&  ...&  ...&  ...& & & \\            
	     & $W_c$&      &1380:&   ...&  107::&  503& ...&  ...&     0:&  ...&  ...&  ...&  ...& & & \\            
\noalign{\vskip 7.5pt}  
F05156$-$3024& $F$&   0.098& 0.82& 0.137:& 0.089& 1.34&  ...&  ...&0.143&0.160&  ...&  ...&  ...& 0.97& 1.47&... \\ 
	     & $EW$&       &     & 11.8:&   7.6&  119&  ...&  ...& 14.7& 17.7&  ...&  ...&  ...& & & \\            
	     & $W$&        &     &  580::&   541:&  715&  ...&  ...&  657:&  787:&  ...&  ...&  ...& & & \\            
	     & $W_c$&      &  950&    0::&     0:&  144&  ...&  ...&    0:&  359:&  ...&  ...&  ...& & & \\            
\noalign{\vskip 7.5pt}  
\tableline
\end{tabular}
\end{center}
\end{table*}

\clearpage

\begin{table*}
\tablenum{3}
\caption{Optical$^a$ and Near-Infrared Line Measurements.} \label{tbl-3}

\begin{center}
\begin{tabular}{llccccccccccccccc}
\tableline
\tableline
IRAS &  & H$\beta_{\rm nl}$ & H$\alpha_{\rm nl}$ & [Fe~II] & H$_2$ & Pa$\alpha_{\rm nl}$ & Pa$\alpha_{\rm bl}$ & 
Br$\delta$ & H$_2$ 1.958 & H$_2$ &He I &H$_2$ & Br$\gamma$ & \multicolumn{3}{c}{E(B--V)$_{\rm nl}$}   \\
\cline{15-17}\\
 & & & & 1.644 & 1.835 & 1.875&  1.875& 1.945 & + [Si~VI] 1.962 & 2.033 & 2.058 & 2.121 &  & H$\alpha$/H$\beta$  & 
 Pa$\alpha$/H$\alpha$ & Br$\gamma$/H$\alpha$\\
(1) & (2) & (3) & (4) & (5) & (6) & (7) & (8) & (9) & (10) & (11) & (12) & (13) & (14) & (15) & (16) & (17)\\
\tableline
\noalign{\vskip 7.5pt}  
%name                   hb    ha  1.644  1.835 pa_nl pa_bl  brd  1.958 2.033 2.058 2.121  brg  ha/hb pa/ha brg/ha  
F05189$-$2524$^b$& $F$&  0.206&  2.9&   ...&  ...& ...& 19.5&  1.76:& 1.72:&  ...&  ...&  ...& 2.2:& 2.03:& ...& ...\\ 
  	         & $EW$&      &     &   ...&  ...& ...& 44.5&  4.0:&  3.9:&  ...&  ...&  ...& 4.8:& & & \\            
	         & $W$&       &     &   ...&  ...& ...& 2711& 1702::& 1702::&  ...&  ...&  ...&2306::& & & \\            
	         & $W_c$&     &  540&   ...&  ...& ...& 2619& 1552::& 1552::&  ...&  ...&  ...&2197::& & & \\            
\noalign{\vskip 7.5pt}  
F08201+2801  & $F$&  0.076& 0.36&   ...&0.063:&0.628&  ...&  ...&0.146&  ...&  ...&0.097::&  ...& 0.53& 1.46&... \\ 
	     & $EW$&      &     &   ...&  8.9:& 90.2&  ...&  ...& 24.8&  ...&  ...&   19::&  ...& & & \\            
	     & $W$&       &     &   ...&  715::&  743&  ...&  ...&  829&  ...&  ...&  ...&  ...& & & \\            
	     & $W_c$&     &  510&   ...&  147::&  250&  ...&  ...&  443&  ...&  ...&  ...&  ...& & & \\            
\noalign{\vskip 7.5pt}  
F08572+3915 NW& $F$& 0.0445& 0.89&   ...& 0.177& 2.33&  ...&0.185:&0.300:&  ...&0.131&0.178&0.232& 1.94& 1.73& 1.75 \\ 
	      & $EW$&      &     &   ...&   9.0&  115&  ...&  8.1:& 12.9:&  ...&  4.4&  5.3&  6.5& & & \\            
	      & $W$&       &     &   ...&   783:&  844&  ...&  749::& 1182::&  ...& 1046:&  664:&  752:& & & \\            
	      & $W_c$&     &  70:&   ...&   351:&  472&  ...&  266::&  952::&  ...&  777:&    0:&  275:& & & \\            
\noalign{\vskip 7.5pt}  
F10091+4704  & $F$& 0.0144& 0.18&0.039:& 0.088&0.218&  ...&  ...&0.119&  ...&  ...&  ...&  ...& 1.36& 1.31&... \\ 
	     & $EW$&      &     &  5.7:&    17.0& 43.9&  ...&  ...& 25.3&  ...&  ...&  ...&  ...& & & \\            
	     & $W$&       &     &  838::&   661:& 1027&  ...&  ...&  569:&  ...&  ...&  ...&  ...& & & \\            
	     & $W_c$&     &  360&  460::&     0:&  752&  ...&  ...&    0:&  ...&  ...&  ...&  ...& & & \\            
\noalign{\vskip 7.5pt}  
F10190+1322  & $F$&  0.118& 1.07&   ...& 0.163:&0.789&  ...&  ...&0.136&0.050:&  ...&0.129:&0.133:& 1.15& 1.00& 1.32: \\ 
	     & $EW$&      &     &   ...&  13.2:& 65.6&  ...&  ...& 12.2&  4.6:&  ...& 13.4:& 14.3:& & & \\            
	     & $W$&       &     &   ...&  1515::&  929&  ...&  ...&  965:&  769::&  ...&  856::&2337::& & & \\            
	     & $W_c$&     &  ...&   ...&  1317::&  549&  ...&  ...&  607:&  170::&  ...&  413::&2213::& & & \\            
\noalign{\vskip 7.5pt}  
F10594+3818  & $F$&  0.212&1.25:&   ...&0.051::& 1.14&  ...&0.073:&0.156&  ...& 0.12::&0.136:&0.28:& 0.75:& 1.11:& 1.63: \\ 
	     & $EW$&      &     &   ...&  8.3::&  198&  ...&   14:& 30.4&  ...&   22::& 28.3:&  64:& & & \\            
	     & $W$&       &     &   ...&  ...&  931&  ...& 1185::& 1001:&  ...&    ...&  633::&1386::& & & \\            
	     & $W_c$&     &  230&   ...&  ...&  552&  ...&  917::&  663:&  ...&    ...&    0::&1165::& & & \\            
\noalign{\vskip 7.5pt}  
F11028+3130  & $F$& 0.0087& 0.03&   ...&   ...&$<$0.041&  ...&  ...&  ...&  ...&  ...&  ...&  ...& 0.10& $<$1.38&... \\ 
	     & $EW$&      &     &   ...&   ...& $<$15.6&  ...&  ...&  ...&  ...&  ...&  ...&  ...& & & \\            
	     & $W$&       &     &   ...&   ...&    ...&  ...&  ...&  ...&  ...&  ...&  ...&  ...& & & \\            
	     & $W_c$&     &  ...&   ...&   ...&    ...&  ...&  ...&  ...&  ...&  ...&  ...&  ...& & & \\            
\noalign{\vskip 7.5pt}  
\tableline
\end{tabular}
\end{center}
\end{table*}

\clearpage

\begin{table*}
\tablenum{3}
\caption{Optical$^a$ and Near-Infrared Line Measurements.} \label{tbl-3}

\begin{center}
\begin{tabular}{llccccccccccccccc}
\tableline
\tableline
IRAS &  & H$\beta_{\rm nl}$ & H$\alpha_{\rm nl}$ & [Fe~II] & H$_2$ & Pa$\alpha_{\rm nl}$ & Pa$\alpha_{\rm bl}$ & 
Br$\delta$ & H$_2$ 1.958 & H$_2$ &He I &H$_2$ & Br$\gamma$ & \multicolumn{3}{c}{E(B--V)$_{\rm nl}$}   \\
\cline{15-17}\\
 & & & & 1.644 & 1.835 & 1.875&  1.875& 1.945 & + [Si~VI] 1.962 & 2.033 & 2.058 & 2.121 &  & H$\alpha$/H$\beta$  & 
 Pa$\alpha$/H$\alpha$ & Br$\gamma$/H$\alpha$\\
(1) & (2) & (3) & (4) & (5) & (6) & (7) & (8) & (9) & (10) & (11) & (12) & (13) & (14) & (15) & (16) & (17)\\
\tableline
\noalign{\vskip 7.5pt}  
%name                   hb    ha  1.644  1.835 pa_nl pa_bl  brd  1.958 2.033 2.058 2.121  brg  ha/hb pa/ha brg/ha  
F11095$-$0237& $F$&  0.109& 0.64&   ...& 0.153& 1.23&  ...&0.069:&0.265&0.094:&0.037:&0.263&0.125:& 0.62& 1.56& 1.60: \\ 
	     & $EW$&      &     &   ...&  25.6&  220&  ...& 13.8:& 53.7& 19.2:&  7.8:& 59.0& 29.0:& & & \\            
	     & $W$&       &     &   ...&   865&  950&  ...&  925::&  818&  823::&  487::&  740& 1122::& & & \\            
	     & $W_c$&     &  320&   ...&   431&  583&  ...&  541::&  325&  338::&    0::&    0&  834::& & & \\            
\noalign{\vskip 7.5pt}  
F11180+1623  & $F$&  0.029&0.29:&0.197::$^c$& 0.051&0.544&  ...&0.037:&0.104&  ...&  ...&0.095:&  ...& 1.21& 1.55:&... \\ 
	     & $EW$&      &     &     36.4::&  12.9&  137&  ...& 10.8:& 31.2&  ...&  ...& 30.2:&  ...& & & \\            
	     & $W$&       &     &        ...&   796:&  936&  ...&  886::&  787&  ...&  ...&  787::&  ...& & & \\            
	     & $W_c$&     &  210&        ...&   267:&  560&  ...&  471::&  239&  ...&  ...&  240::&  ...& & & \\            
\noalign{\vskip 7.5pt}  
F11223$-$1244& $F$& 0.004:& 0.20& 0.108& 0.118&0.854&  ...&  ...&0.213&0.084:&  ...&  ...&  ...& 2.68& 1.99&... \\ 
	     & $EW$&      &     &   9.4&  11.1& 81.6&  ...&  ...& 21.5&  8.5:&  ...&  ...&  ...& & & \\            
	     & $W$&       &     &   672:&   635:&  695&  ...&  ...&  729&  561::&  ...&  ...&  ...& & & \\            
	     & $W_c$&     &  960&     0:&     0:&    0&  ...&  ...&  204&    0::&  ...&  ...&  ...& & & \\            
\noalign{\vskip 7.5pt}  
F11506+1331  & $F$&  0.047& 0.47&   ...&   ...& 3.26&  ...&0.186:&0.181:&  ...&0.175:&0.241:&0.278:& 1.13& 2.21& 2.13:\\ 
	     & $EW$&      &     &   ...&   ...&  215&  ...& 12.9:& 12.6:&  ...& 12.5:& 17.6:& 20.7:& & & \\            
	     & $W$&       &     &   ...&   ...&  916&  ...& 1149::&  763::&  ...&  718::&906::&901::& & & \\            
	     & $W_c$&     &  80:&   ...&   ...&  526&  ...&  871::&  138::&  ...&    0::&507::&500::& & & \\            
\noalign{\vskip 7.5pt}  
%name                   hb    ha  1.644  1.835 pa_nl pa_bl  brd  1.958 2.033 2.058 2.121  brg  ha/hb pa/ha brg/ha  
F11582+3020  & $F$&  0.040& 0.25&   ...&0.070::&0.560&  ...&  ...&0.120:&  ...&  ...&  ...&  ...& 0.76& 1.64&... \\ 
	     & $EW$&      &     &   ...& 10.8::& 90.6&  ...&  ...& 20.0:&  ...&  ...&  ...&  ...& & & \\            
	     & $W$&       &     &   ...&   ...&  676&  ...&  ...&  969::&  ...&  ...&  ...&  ...& & & \\            
	     & $W_c$&     &  250&   ...&   ...&    0&  ...&  ...&  670::&  ...&  ...&  ...&  ...& & & \\            
\noalign{\vskip 7.5pt}  
F12018+1941  & $F$& ...&0.182$^d$& ...& 0.088:&0.785&  ...&  ...&0.183&0.041::&  ...&0.290&  ...& ...& 2.00&... \\ 
	     & $EW$&      &     & ...&  12.9:&  120&  ...&  ...& 30.6&  6.9::&  ...& 52.6&  ...& & & \\            
	     & $W$&       &     & ...&   783::&  841&  ...&  ...&  886&   ...&  ...&  996:&  ...& & & \\            
	     & $W_c$&     &  ...& ...&   224::&  381&  ...&  ...&  471&   ...&  ...&  655:&  ...& & & \\            
\noalign{\vskip 7.5pt}  
F12032+1707  & $F$&  0.214& 1.65& 0.348& 0.313&0.981&  ...&0.193&0.550&0.338&  ...&  ...&  ...& 0.94& 0.93&... \\ 
	     & $EW$&      &     &  54.6&  63.0&  205&  ...& 46.1&  133& 92.1&  ...&  ...&  ...& & & \\            
	     & $W$&       &     &   824&   924&  851&  ...& 1334:&  817& 1442:&  ...&  ...&  ...& & & \\            
	     & $W_c$&     &  580&   342&   540&  402&  ...& 1103:&  323& 1232:&  ...&  ...&  ...& & & \\            
\noalign{\vskip 7.5pt}  
\tableline
\end{tabular}
\end{center}
\end{table*}

\clearpage

\begin{table*}
\tablenum{3}
\caption{Optical$^a$ and Near-Infrared Line Measurements.} \label{tbl-3}

\begin{center}
\begin{tabular}{llccccccccccccccc}
\tableline
\tableline
IRAS &  & H$\beta_{\rm nl}$ & H$\alpha_{\rm nl}$ & [Fe~II] & H$_2$ & Pa$\alpha_{\rm nl}$ & Pa$\alpha_{\rm bl}$ & 
Br$\delta$ & H$_2$ 1.958 & H$_2$ &He I &H$_2$ & Br$\gamma$ & \multicolumn{3}{c}{E(B--V)$_{\rm nl}$}   \\
\cline{15-17}\\
 & & & & 1.644 & 1.835 & 1.875&  1.875& 1.945 & + [Si~VI] 1.962 & 2.033 & 2.058 & 2.121 &  & H$\alpha$/H$\beta$  & 
 Pa$\alpha$/H$\alpha$ & Br$\gamma$/H$\alpha$\\
(1) & (2) & (3) & (4) & (5) & (6) & (7) & (8) & (9) & (10) & (11) & (12) & (13) & (14) & (15) & (16) & (17)\\
\tableline
\noalign{\vskip 7.5pt}  
F12112+0305  & $F$&  0.104& 0.80&   ...&  ...& 2.01&  ...&0.085:&0.144:&0.098:&0.093:&0.141:&0.217:& 0.92& 1.70& 1.77: \\ 
	     & $EW$&      &     &   ...&  ...&  118&  ...&  5.2:&  9.1:&  6.6:&  6.4:& 10.1:& 15.7:& & & \\            
	     & $W$&       &     &   ...&  ...&  896&  ...&  686::&  725::&  828::&  458::&  630::&  670::& & & \\          
	     & $W_c$&     &  ...&   ...&  ...&  559&  ...&    0::&  191::&  441::&    0::&    0::&    0::& & & \\           
\noalign{\vskip 7.5pt}  
F12447+3721  & $F$&  0.391&2.06:& 0.161&   ...& 1.62& ...&0.130:&0.072:&  ...&  ...&  ...&0.222:& 0.61& 1.03:& 1.25: \\ 
	     & $EW$&      &     &  41.4&   ...&  543& ...& 54.2:& 29.7:&  ...&  ...&  ...& 84.0:& & & \\            
	     & $W$&       &     &   802:&  ...&  727& ...&1055::&  632::&  ...&  ...&  ...& 785::& & & \\            
	     & $W_c$&     &  190&   283:&  ...&    0& ...& 742::&    0::&  ...&  ...&  ...& 232::& & & \\            
\noalign{\vskip 7.5pt}  
%name                   hb    ha  1.644  1.835 pa_nl pa_bl  brd  1.958 2.033 2.058 2.121  brg  ha/hb pa/ha brg/ha  
F13106$-$0922& $F$& 0.0495& 0.33&   ...&0.055:&0.323&  ...&0.035:&0.122&0.091&  ...&0.211:&  ...& 0.76& 1.20&... \\ 
	     & $EW$&      &     &   ...& 12.2:& 75.8&  ...&  9.1:& 32.3& 26.9&  ...& 71.2:&  ...& & & \\            
	     & $W$&       &     &   ...&  801::&  874&  ...&  790::&  712&  854:&  ...& 1201::&  ...& & & \\            
	     & $W_c$&     &  160&   ...&  281::&  448&  ...&  250::&    0&  407:&  ...&  938::&  ...& & & \\            
\noalign{\vskip 7.5pt}  
F13305$-$1739& $F$&  0.955&9.55:&   ...& 0.264& 1.73& 1.48&0.249:&0.973&0.191&  ...&0.217:&0.155::& 1.18:& 0.28:& 0.31:: \\ 
	     & $EW$&      &     &   ...&  18.8&  129&  110& 19.7:& 79.0& 16.5&  ...& 19.8:& 15.0::& & & \\            
	     & $W$&       &     &   ...&  1030:& 1072& 2991& 1813::& 1679&  965:&  ...&  616::& ...& & & \\            
	     & $W_c$&     & 1200&   ...&   707:&  766& 2896& 1651::& 1503&  607:&  ...&    0::& ...& & & \\            
\noalign{\vskip 7.5pt}  
F13428+5608$^e$  & $F$&  0.821& 8.21&   ...& 0.847:& 8.84&  ...&0.700:& 2.02&0.754& ...& 1.42&0.822& 1.22& 1.25&... \\ 
	         & $EW$&      &     &   ...&  12.8:&  132&  ...& 11.2:& 32.9& 12.7& ...& 24.9& 14.7& & & \\            
	         & $W$&       &     &   ...&   976::& 1168&  ...& 1171::& 1167& 1153:& ...& 1078& 1034:& & & \\            
	         & $W_c$&     &  480&   ...&   625::&  895&  ...&  899::&  894&  876:& ...&  774&  712:& & & \\            
\noalign{\vskip 7.5pt}  
F13443+0802 NE& $F$&  0.101& 1.01&   ...&0.098:& 1.00&  ...&  ...&0.167&  ...&  ...&  ...&  ...& 1.22& 1.16&... \\ 
	      & $EW$&      &     &   ...&  5.8:& 60.2&  ...&  ...& 11.4&  ...&  ...&  ...&  ...& & & \\            
	      & $W$&       &     &   ...&  810::&  954&  ...&  ...&1038:&  ...&  ...&  ...&  ...& & & \\            
	      & $W_c$&     &   500&   ...&  307::&  590&  ...&  ...& 717:&  ...&  ...&  ...&  ...& & & \\            
\noalign{\vskip 7.5pt}  
F13443+0802 SW& $F$&  0.025& 0.36&   ...&   ...&0.211&  ...&  ...&0.133&  ...&  ...&0.104:&  ...& 1.57& 0.92&... \\ 
	      & $EW$&      &     &   ...&   ...& 59.6&  ...&  ...& 40.5&  ...&  ...& 35.0:&  ...& & & \\            
	      & $W$&       &     &   ...&   ...&  943&  ...&  ...&1455:&  ...&  ...&1094::&  ...& & & \\            
	      & $W_c$&     &  410&   ...&   ...&  572&  ...&  ...&1246:&  ...&  ...& 797::&  ...& & & \\            
\noalign{\vskip 7.5pt}  
\tableline
\end{tabular}
\end{center}
\end{table*}

\clearpage

\begin{table*}
\tablenum{3}
\caption{Optical$^a$ and Near-Infrared Line Measurements.} \label{tbl-3}

\begin{center}
\begin{tabular}{llccccccccccccccc}
\tableline
\tableline
IRAS &  & H$\beta_{\rm nl}$ & H$\alpha_{\rm nl}$ & [Fe~II] & H$_2$ & Pa$\alpha_{\rm nl}$ & Pa$\alpha_{\rm bl}$ & 
Br$\delta$ & H$_2$ 1.958 & H$_2$ &He I &H$_2$ & Br$\gamma$ & \multicolumn{3}{c}{E(B--V)$_{\rm nl}$}   \\
\cline{15-17}\\
 & & & & 1.644 & 1.835 & 1.875&  1.875& 1.945 & + [Si~VI] 1.962 & 2.033 & 2.058 & 2.121 &  & H$\alpha$/H$\beta$  & 
 Pa$\alpha$/H$\alpha$ & Br$\gamma$/H$\alpha$\\
(1) & (2) & (3) & (4) & (5) & (6) & (7) & (8) & (9) & (10) & (11) & (12) & (13) & (14) & (15) & (16) & (17)\\
\tableline
\noalign{\vskip 7.5pt}  
%name                   hb    ha  1.644  1.835 pa_nl pa_bl  brd  1.958 2.033 2.058 2.121  brg  ha/hb pa/ha brg/ha  
F13454$-$2956& $F$& 0.0084& 0.12&   ...& 0.086:& 1.28&  ...&0.064:&0.266&  ...&  ...&0.164:&0.076:& 1.54& 2.49& 2.21: \\ 
	     & $EW$&      &     &   ...&   7.5:&  114&  ...&  6.1:& 25.4&  ...&  ...& 16.8:&  8.0:& & & \\            
	     & $W$&       &     &   ...&  728::&  917&  ...&  676::&  949&  ...&  ...& 842::&  523::& & & \\            
	     & $W_c$&     &  70:&   ...&    0::&  528&  ...&    0::&  582&  ...&  ...& 383::&    0::& & & \\            
\noalign{\vskip 7.5pt}  
F13509+0442  & $F$&  0.125& 1.14&   ...&  ...& 2.41&  ...&0.190&0.146&  ...&0.188:&0.232:&0.192:& 1.20& 1.57& 1.48: \\ 
	     & $EW$&      &     &   ...&  ...&  252&  ...& 21.9& 17.2&  ...& 23.2:& 30.6:& 26.5:& & & \\            
	     & $W$&       &     &   ...&  ...&  875&  ...&  842:&  803:&  ...& 1063::& 1002::&  700::& & & \\            
	     & $W_c$&     &  ...&   ...&  ...&  450&  ...&  384:&  287:&  ...&  754::&  665::&    0::& & & \\            
\noalign{\vskip 7.5pt}  
F14053$-$1958& $F$& 0.026:&0.11:&   ...&   ...&0.225&  ...&  ...&   ...&  ...&  ...&  ...&  ...& 0.31& 1.59:&... \\ 
	     & $EW$&      &     &   ...&   ...& 71.5&  ...&  ...&   ...&  ...&  ...&  ...&  ...& & & \\            
	     & $W$&       &     &   ...&   ...&  848&  ...&  ...&   ...&  ...&  ...&  ...&  ...& & & \\            
	     & $W_c$&     &  420&   ...&   ...&  397&  ...&  ...&   ...&  ...&  ...&  ...&  ...& & & \\            
\noalign{\vskip 7.5pt}  
F14070+0525  & $F$& 0.0044& 0.11& 0.052:&   ...&0.236&  ...&  ...&0.131:&  ...&  ...&  ...&  ...& 2.08& 1.62&... \\ 
	     & $EW$&      &     &   9.8:&   ...& 55.5&  ...&  ...& 35.9:&  ...&  ...&  ...&  ...& & & \\            
	     & $W$&       &     &  780::&   ...&  702&  ...&  ...&  813::&  ...&  ...&  ...&  ...& & & \\            
	     & $W_c$&     &  ...&  212::&   ...&    0&  ...&  ...&  315::&  ...&  ...&  ...&  ...& & & \\            
\noalign{\vskip 7.5pt}  
F14394+5332  & $F$&  0.584& 4.49&   ...& 0.585& 3.97&  ...&0.391& 1.00&0.273&  ...&0.840&0.659:& 0.93& 1.14& 1.45 \\ 
	     & $EW$&      &     &   ...&  25.9&  184&  ...& 19.0& 49.7& 13.7&  ...& 47.0& 38.4:& & & \\            
	     & $W$&       &     &   ...&  1060& 1031&  ...& 1336:& 1026&  760:&  ...&  810& 2319::& & & \\            
	     & $W_c$&     & 1870&   ...&   750&  708&  ...& 1106:&  700&  120:&  ...&  305& 2195::& & & \\            
\noalign{\vskip 7.5pt}  
F15250+3609  & $F$&  0.285&  1.9&   ...&0.201:& 3.13&  ...&0.240:&0.413&0.154:&0.241::&0.432&0.290& 0.76& 1.48& 1.47 \\ 
	     & $EW$&      &     &   ...&  6.4:&  103&  ...&  8.6:& 15.0&  6.0:&  9.6::& 18.0& 12.5& & & \\            
	     & $W$&       &     &   ...&  914::&  937&  ...&1089::&  845& 735::& ...&  732&  761& & & \\            
	     & $W_c$&     &  150&   ...&  523::&  562&  ...& 790::&  389&   0::& ...&    0&  131& & & \\            
\noalign{\vskip 7.5pt}  
F16156+0146  & $F$&  0.125& 0.78&   ...& 0.110:& 1.15&  ...&0.104::&0.226&  ...&  ...&0.194&0.054::& 0.69& 1.42& 1.06\\ 
	     & $EW$&      &     &   ...&  20.3:&  213&  ...& 19.8::& 43.2&  ...&  ...& 35.6&  9.7::& & & \\            
	     & $W$&       &     &   ...&   961::& 1190&  ...& ...& 1056&  ...&  ...&  706&   ...& & & \\            
	     & $W_c$&     &  590&   ...&   600::&  925&  ...& ...&  743&  ...&  ...&    0&   ...& & & \\            
\noalign{\vskip 7.5pt}  
\tableline
\end{tabular}
\end{center}
\end{table*}

\clearpage

\begin{table*}
\tablenum{3}
\caption{Optical$^a$ and Near-Infrared Line Measurements.} \label{tbl-3}

\begin{center}
\begin{tabular}{llccccccccccccccc}
\tableline
\tableline
IRAS &  & H$\beta_{\rm nl}$ & H$\alpha_{\rm nl}$ & [Fe~II] & H$_2$ & Pa$\alpha_{\rm nl}$ & Pa$\alpha_{\rm bl}$ & 
Br$\delta$ & H$_2$ 1.958 & H$_2$ &He I &H$_2$ & Br$\gamma$ & \multicolumn{3}{c}{E(B--V)$_{\rm nl}$}   \\
\cline{15-17}\\
 & & & & 1.644 & 1.835 & 1.875&  1.875& 1.945 & + [Si~VI] 1.962 & 2.033 & 2.058 & 2.121 &  & H$\alpha$/H$\beta$  & 
 Pa$\alpha$/H$\alpha$ & Br$\gamma$/H$\alpha$\\
(1) & (2) & (3) & (4) & (5) & (6) & (7) & (8) & (9) & (10) & (11) & (12) & (13) & (14) & (15) & (16) & (17)\\
\tableline
\noalign{\vskip 7.5pt}  
%name                   hb    ha  1.644  1.835 pa_nl pa_bl  brd  1.958 2.033 2.058 2.121  brg  ha/hb pa/ha brg/ha  
F16300+1558  & $F$&  0.026& 0.22&0.064:&   ...&0.560&  ...&  ...&0.119:&  ...&  ...&  ...&  ...& 1.00& 1.71&... \\ 
	     & $EW$&      &     &  9.0:&   ...& 93.8&  ...&  ...& 22.2:&  ...&  ...&  ...&  ...& & & \\            
	     & $W$&       &     &1027::&   ...&  928&  ...&  ...&  959::&  ...&  ...&  ...&  ...& & & \\            
	     & $W_c$&     &  120& 701::&   ...&  547&  ...&  ...&  597::&  ...&  ...&  ...&  ...& & & \\            
\noalign{\vskip 7.5pt}  
F17208$-$0014& $F$&  0.086&  1.6&   ...&   ...& 6.21&  ...&0.692::& 1.05&0.564:&0.285:& 1.14&0.780& 1.87& 1.89& 2.03 \\ 
	     & $EW$&      &     &   ...&   ...& 90.7&  ...& 10.5::& 16.2&  9.0:&  4.6:& 19.2& 13.5& & & \\            
	     & $W$&       &     &   ...&   ...&  918&  ...&    ...&  806& 932::&  798::&759&  794& & & \\            
	     & $W_c$&     &  ...&   ...&   ...&  529&  ...&    ...&  294& 553::&  273::&114&  260& & & \\            
\noalign{\vskip 7.5pt}  
F22491$-$1808& $F$&  0.368&  2.3&   ...&0.049:&0.719&  ...&  ...&0.110:&  ...&  ...&  ...& ...& 0.81& 0.53& ... \\ 
	     & $EW$&      &     &   ...&  4.2:& 64.8&  ...&  ...&  10.9:&  ...&  ...&  ...&...& & & \\            
	     & $W$&       &     &   ...&  565::&  882&  ...&  ...&1138::&  ...&  ...&  ...&...& & & \\            
	     & $W_c$&     &  ...&   ...&    0::&  536&  ...&  ...& 898::&  ...&  ...&  ...&...& & & \\            
\noalign{\vskip 7.5pt}  
F23365+3604  & $F$&  0.208&  1.6&   ...& 0.372:& 4.94&  ...&  ...&0.468&  ...&  ...&0.468&0.357& 0.88& 1.82& 1.67 \\ 
	     & $EW$&      &     &   ...&  9.9:&  134&  ...&  ...& 13.5&  ...&  ...& 15.3& 12.0& & & \\            
	     & $W$&       &     &   ...&  874::&  927&  ...&  ...&  777:&  ...&  ...&  800:&  683:& & & \\            
	     & $W_c$&     &  260&   ...&  524::&  607&  ...&  ...&  337:&  ...&  ...&  388:&    0:& & & \\            
\noalign{\vskip 22.5pt}  
F11058$-$1131 (LIG)& $F$&2.64$^f$&13.9$^f$&   ...&0.302:& 2.27& 6.94&  ...&0.585&  ...&  ...&0.14::&0.14::& 0.49& 0.83& 0.07:: \\ 
	     & $EW$&      &       &   ...&  3.2:&  24.3& 74.3&  ...& 0.64&  ...&  ...& 1.6::& 1.6::& & & \\            
	     & $W$&       &       &   ...& 1097::&  903& 7218&  ...&  934&  ...&  ...& 835::& 684::& & & \\            
	     & $W_c$&     &220$^f$&   ...&  801::&  503& 7179&  ...&  557&  ...&  ...& 368::&   0::& & & \\            
%name                   hb    ha  1.644  1.835 pa_nl pa_bl  brd  1.958 2.033 2.058 2.121  brg  ha/hb pa/ha brg/ha  
\noalign{\vskip 7.5pt}  
\tableline
\end{tabular}
\end{center}
\end{table*}

\clearpage

\begin{table*}
\tablecomments{Meaning of columns and rows:}

\tablenotetext{}{Column (1) -- Name of object.}

\tablenotetext{}{Column (2) -- Meaning of rows. We
list the observed line flux (10$^{-14}$ ergs s$^{-1}$ cm$^{-2}$) on
the first row, the observed equivalent width (\AA) on the second row,
the observed line width (km~s$^{-1}$) on the third row, and the line
width corrected for the instrumental resolution (km~s$^{-1}$) on the
fourth row.}

\tablenotetext{}{Columns (3) -- (14) -- Line fluxes, equivalent 
widths, and line widths for narrow H$\beta$, narrow H$\alpha$, [Fe~II]
$\lambda$1.644, H$_2$ $\lambda$1.835, narrow Pa$\alpha$, broad
Pa$\alpha$, Br$\delta$, the blend of H$_2$ $\lambda$1.958 and [Si~VI]
$\lambda$1.962, H$_2$ $\lambda$2.033, He~I $\lambda$2.058, H$_2$
$\lambda$2.122, and narrow Br$\gamma$, respectively.}

\tablenotetext{}{Column (15) -- (17) -- 
The color excesses derived from H$\alpha$/H$\beta$ were
taken directly from Kim et al. (1998) or Veilleux et al. (1999).
The infrared color excesses were derived using
$$ E(B-V) {\rm (line~2/line~1)} = \frac{2.5}{\frac{A_1}{E(B-V)}+\frac{A_2}{E(B-V)}}~log~\left[\frac{\left(\frac{F_2}{F_1}\right)}{\left(\frac{F_2}{F_1}\right)_0}\right] $$
with the intrinsic flux ratios and interstellar extinction coefficients listed
in Table 2 of Veilleux et al. (1997a). 
The reddenings enclosed in parentheses were calculated using the sum of 
the broad and narrow-line fluxes; they should therefore be considered 
upper limits to the actual values.}

\tablenotetext{^a}{Unless otherwise noted, all of the optical measurements are
from Kim et al. (1995, 1998) or Veilleux et al. (1995, 1999). }

\tablenotetext{^b}{Observed fluxes (10$^{-14}$ erg~s$^{-1}$~cm$^{-2}$), 
equivalent widths (\AA), observed line widths (km~s$^{-1}$), and corrected line widths (km~s$^{-1}$)
for other lines:
He~I$_{\rm nl}$ $\lambda$1.083 = 1.99, 6.6, 753, 456;  
He~I$_{\rm bl}$ $\lambda$1.083 = 11.3:, 38:, 6171:, and 6142:; 
Pa$\beta_{\rm nl}$ = 7.34, 20.0, 1687, and 1577; 
Pa$\beta_{\rm bl}$ = 3.2:, 8.8:, 8936:, and 8916:; 
The color excess derived from Pa$\beta_{\rm nl}$/H$\alpha_{\rm nl}$ is 2.60.}

\tablenotetext{^c}{Line affected by atmospheric correction?}

\tablenotetext{^d}{From Armus et al. (1989)}

\tablenotetext{^e}{Observed fluxes (10$^{-14}$ erg~s$^{-1}$~cm$^{-2}$), 
equivalent widths (\AA), observed line widths (km~s$^{-1}$), and
corrected line widths (km~s$^{-1}$) for other lines: He~I$_{\rm nl}$
$\lambda$1.083 = 3.21, 49.9, 1108, and 897; [Fe~II] $\lambda$1.256 =
1.50, 45.0, 939, and 677; Pa$\beta_{\rm nl}$ = 1.35, 40.0, 718, and
306; The color excess derived from Pa$\beta_{\rm nl}$/H$\alpha_{\rm
nl}$ is 0.77.}

\tablenotetext{^f}{From Osterbrock \& De Robertis (1985). The corrected line 
width refers to narrow H$\alpha$ rather than [O~III] $\lambda$0.5007.}

\end{table*}

\clearpage

%\end{document}

\clearpage

\begin{figure}
%\epsscale{1.2}
%\plotone{fig1_1.ps}
\caption{ Reduced near-infrared spectra of the {\em IRAS}
ultraluminous infrared galaxies in the new sample.  $f_\lambda$ is plotted
versus $\lambda_{\rm observed}$.  The units of the vertical axis are
$10^{-11}$~ergs~s$^{-1}$~cm$^{-2}$~\micron$^{-1}$, while the
wavelength scale is in \micron.}
\end{figure}

%\setcounter{figure}{0}
%\begin{figure}
%%\epsscale{1.2}
%%\plotone{fig1_2.ps}
%\caption{ (cont'd)}
%\end{figure}

%\setcounter{figure}{0}
%\begin{figure}
%%\epsscale{1.2}
%%\plotone{fig1_3.ps}
%\caption{ (cont'd)}
%\end{figure}

%\setcounter{figure}{0}
%\begin{figure}
%%\epsscale{1.2}
%%\plotone{fig1_4.ps}
%\caption{ (cont'd)}
%\end{figure}

%\setcounter{figure}{0}
%\begin{figure}
%%\epsscale{1.2}
%%\plotone{fig1_5.ps}
%\caption{ (cont'd)}
%\end{figure}

%\setcounter{figure}{0}
%\begin{figure}
%%\epsscale{1.2}
%%\plotone{fig1_6.ps}
%\caption{ (cont'd)}
%\end{figure}

%\setcounter{figure}{1}
\begin{figure}
%\epsscale{1.2}
%\plotone{fig2.ps}
\caption{ Distributions of the (a) H$_2$ $\lambda$1.958/Pa$\alpha_{nl}$
and (b) H$_2$ $\lambda$2.122/Pa$\alpha_{nl}$ flux ratios for the
optically classified H~II galaxies (dotted line), LINERs (dashed
line), and Seyfert 2 galaxies (solid line) of the combined sample.}
\end{figure}

\begin{figure}
%\epsscale{1.2}
%\plotone{fig3.ps}
\caption{ Luminosity of Pa$\alpha$ as a function of the infrared
luminosity in units of the solar bolometric luminosity, 3.83 $\times$
10$^{33}$~erg~s$^{-1}$.  The vertical axis of panels (a) and (b) shows
the luminosities in the narrow component of Pa$\alpha$ while the
luminosities plotted in panels (c) and (d) refer to the sum of the
narrow and broad components. Dereddening of the emission-line
luminosities in the left and right panels was carried out using the
color excesses derived from the optical and infrared line ratios,
respectively. Ultraluminous infrared galaxies with obscured BLRs
detected in our near-infrared spectra are represented by filled boxes,
Seyfert~2 galaxies by filled circles, LINERs by open circles, and H~II
galaxies by asterisks.  The filled triangle is the luminous infrared
galaxy F11058-1131. The solid line represents the relation found by
Goldader et al. (1995, 1997a,b) for lower luminosity infrared galaxies
after assuming a Pa$\alpha$/Br$\gamma$ flux ratio of 12 (case B
recombination).  Several data points fall below the solid line,
therefore suggesting a Pa$\alpha$ emission deficit in many of these
objects.}
\end{figure}

\begin{figure}
%\epsscale{1.2}
%\plotone{fig4.ps}
\caption{ Ratios of the dereddened Pa$\alpha$ and infrared
luminosities as a function of the {\em IRAS} $f_{25}/f_{60}$ ratio.
Dereddening was carried out using (a) the optical line ratios and (b)
the infrared line ratios. Meaning of the symbols is the same as in
Fig. 3. The P$\alpha$-to-IR luminosity ratio is on average larger in
those objects with `warm' infrared colors ($f_{25}/f_{60} > 0.2$).}
\end{figure}

\begin{figure}
%\epsscale{1.2}
%\plotone{fig5.ps}
\caption{ Comparison of the H$\beta$ luminosities of optical and
obscured BLRs in infrared galaxies and optically identified QSOs as a
function of their bolometric luminosities. $L_{\rm bol}$ for the QSOs
was determined using the bolometric correction factor 11.8
(i.e. $L_{\rm bol} = 11.8 \nu_B L_\nu(B)$: Elvis et al. 1994; Sanders
\& Mirabel 1996), except for those few sources that were detected by
{\em IRAS}, in which case $L_{\rm bol}$ was taken from Sanders et
al. (1989).  For the ULIGs $L_{\rm bol}$ was taken to be $1.15 \times
L_{\rm ir}$ (Kim \& Sanders 1998).  The H$\beta$ data for the
optically selected QSOs are from Yee (1980) corrected for $H_{\rm o} =
75$~km~s$^{-1}$~Mpc$^{-1}$ and $q_{\rm o} = 0$.  The broad H$\beta$
luminosities of the infrared galaxies were calculated from the
measured broad Pa$\alpha$ fluxes assuming case B recombination (except
for Mrk~463E where the broad Pa$\beta$ flux was used).  The reddening
correction was carried out using the color excesses derived from the
infrared line ratios. Most of the infrared galaxies fall close to the
quasar relation (solid line).}
\end{figure}

\clearpage

\setcounter{figure}{0}
\begin{figure}
%\epsscale{1.2}
\plotone{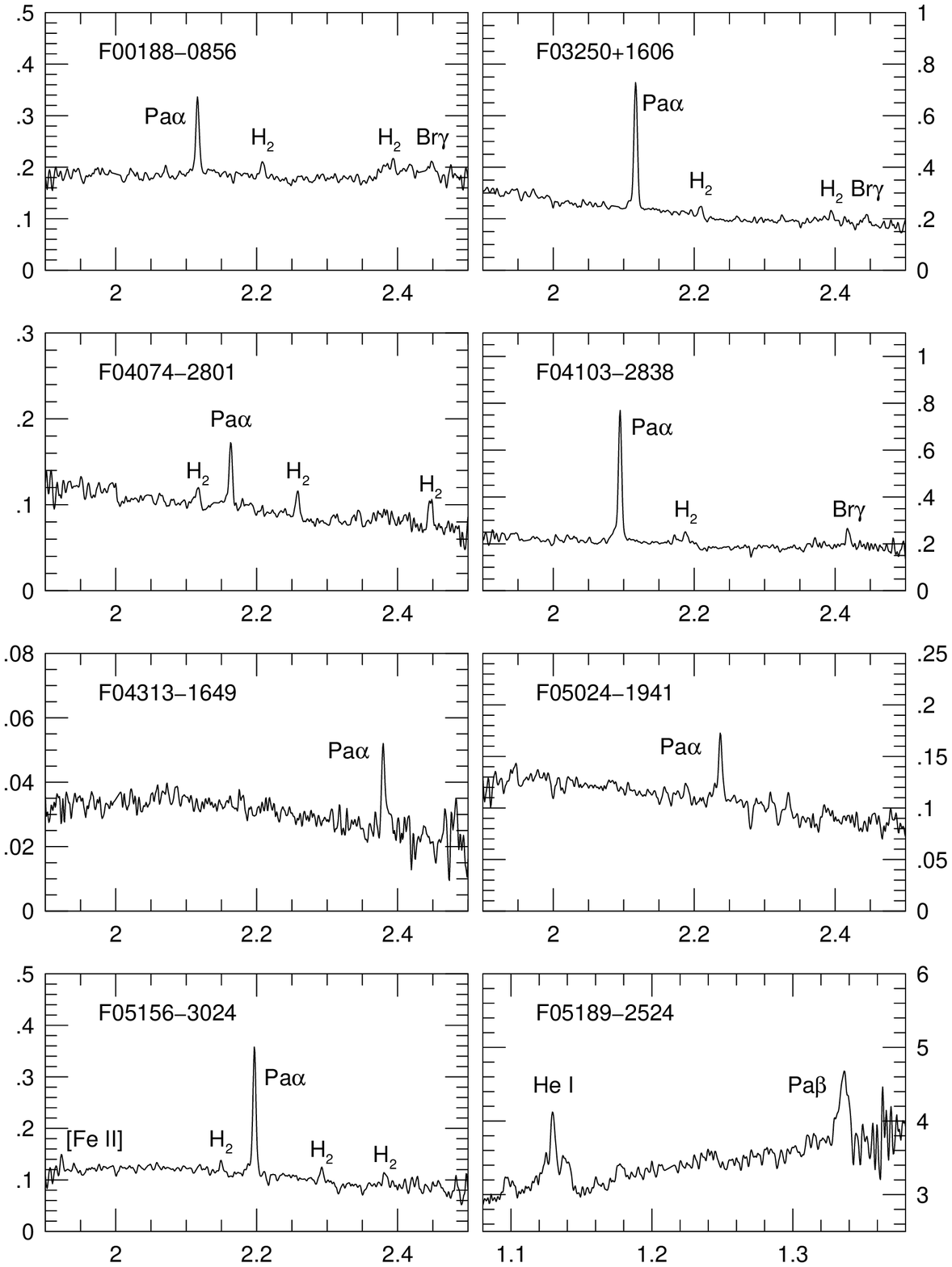}
\caption{ }
\end{figure}

\setcounter{figure}{0}
\begin{figure}
%\epsscale{1.2}
\plotone{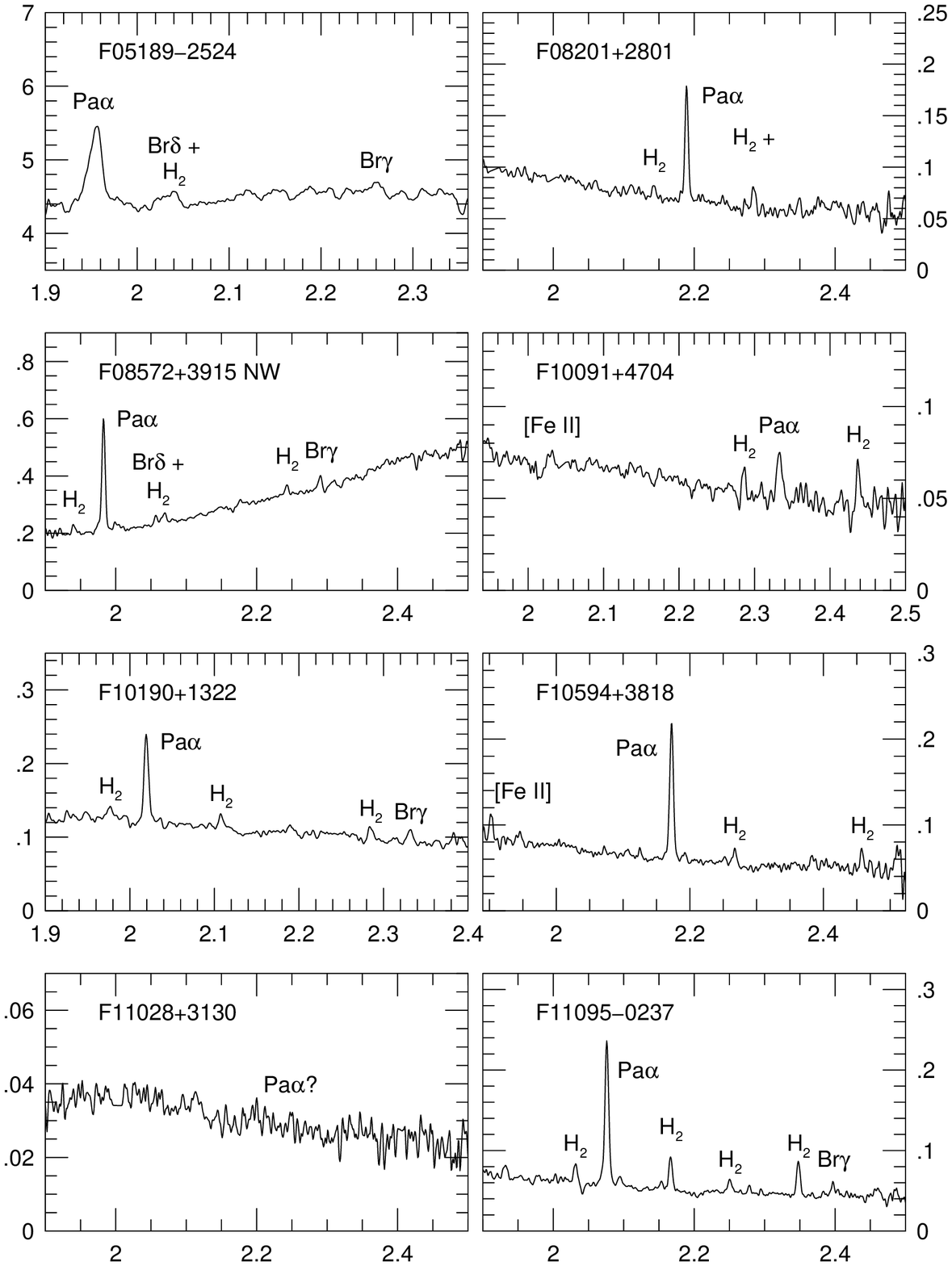}
\caption{ }
\end{figure}

\setcounter{figure}{0}
\begin{figure}
%\epsscale{1.2}
\plotone{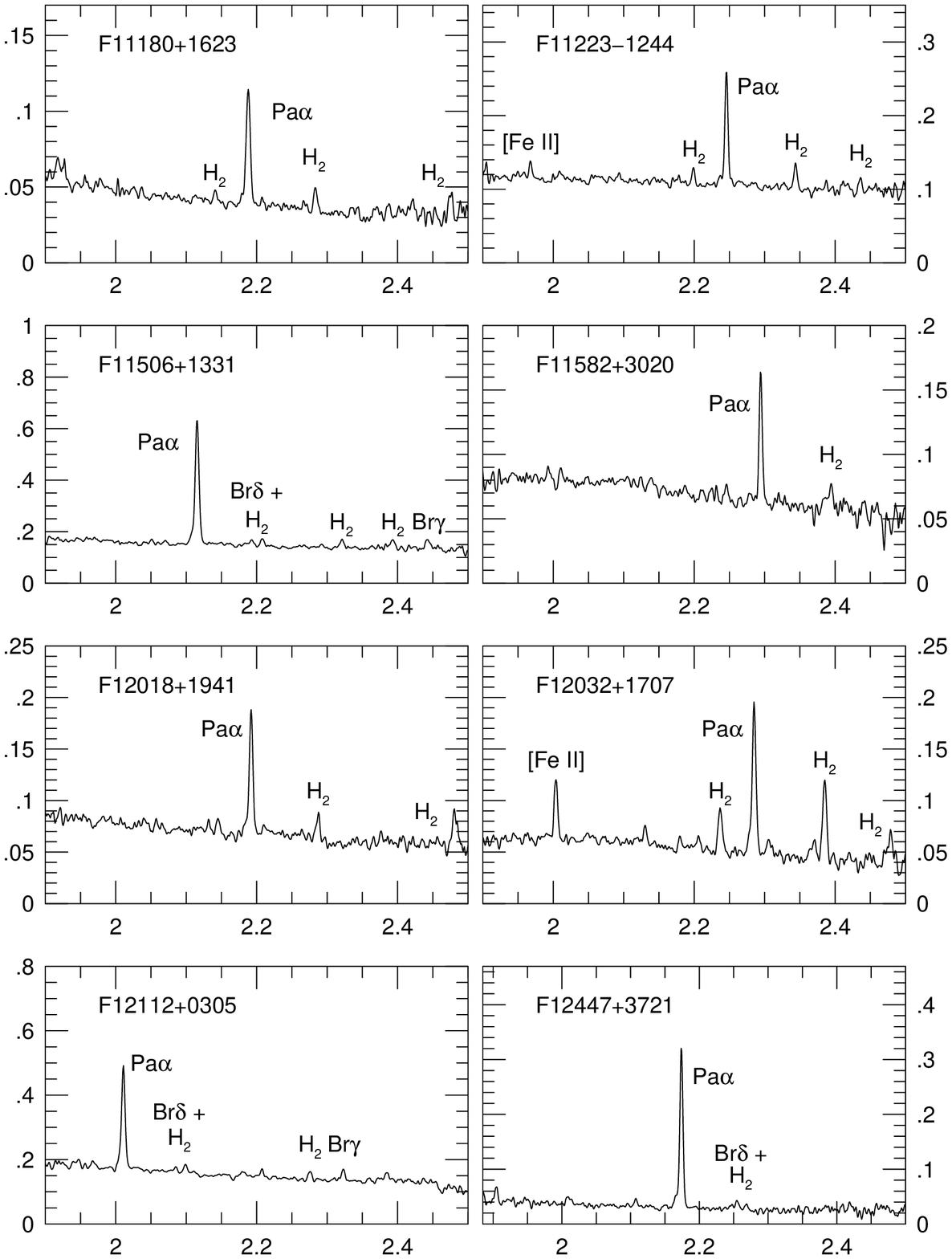}
\caption{ }
\end{figure}

\setcounter{figure}{0}
\begin{figure}
%\epsscale{1.2}
\plotone{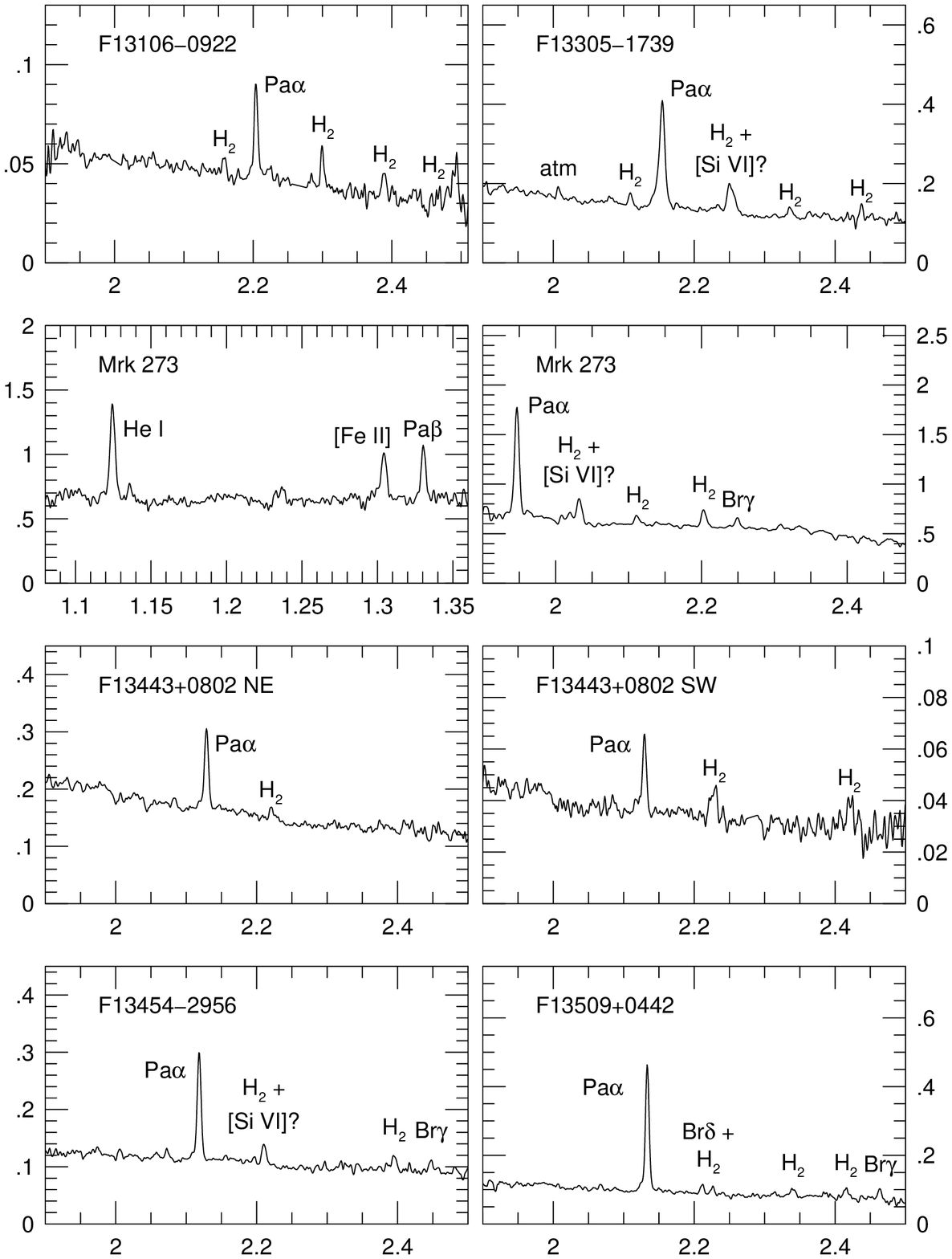}
\caption{ }
\end{figure}

\setcounter{figure}{0}
\begin{figure}
%\epsscale{1.2}
\plotone{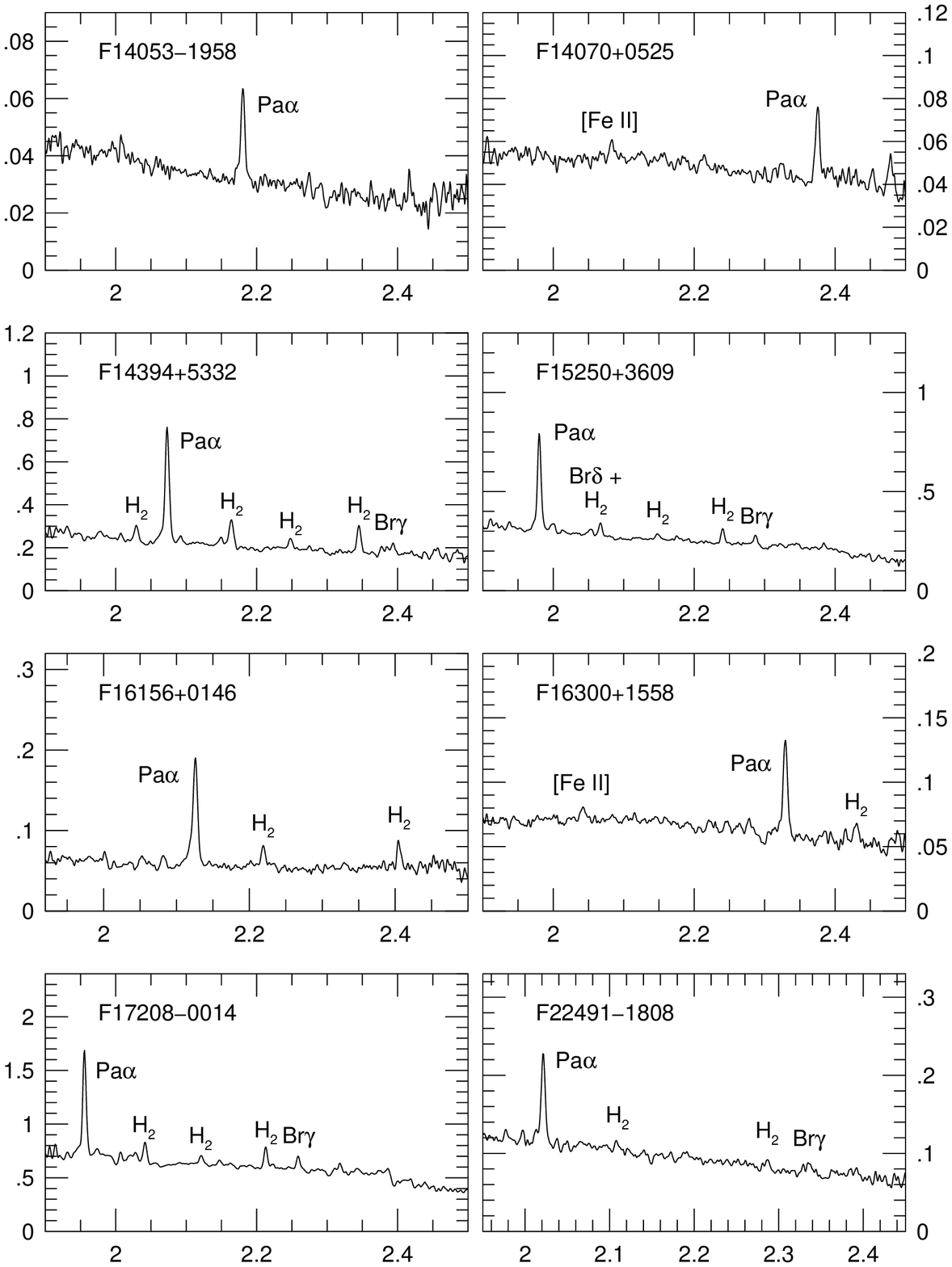}
\caption{ }
\end{figure}

\setcounter{figure}{0}
\begin{figure}
%\epsscale{1.2}
\plotone{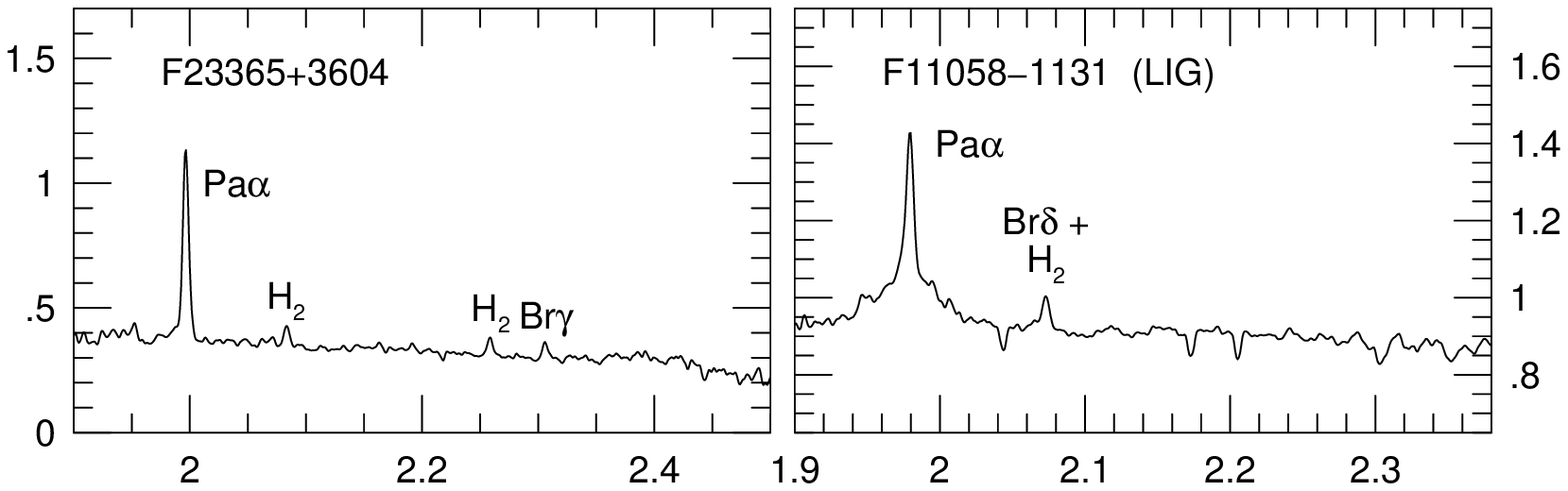}
\caption{ }
\end{figure}

\setcounter{figure}{1}
\begin{figure}
%\epsscale{1.2}
\plotone{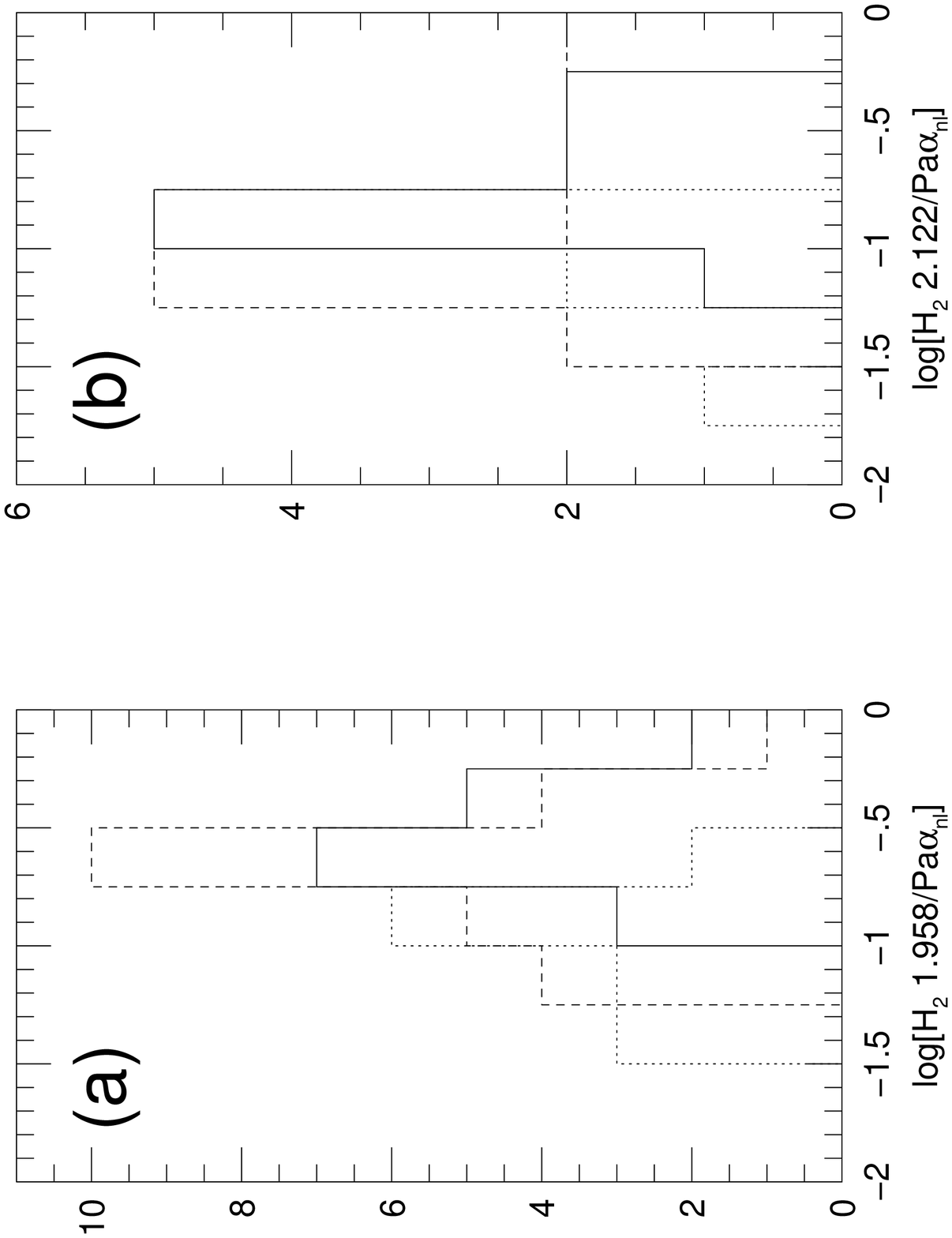}
\caption{ }
\end{figure}

\begin{figure}
%\epsscale{1.2}
\plotone{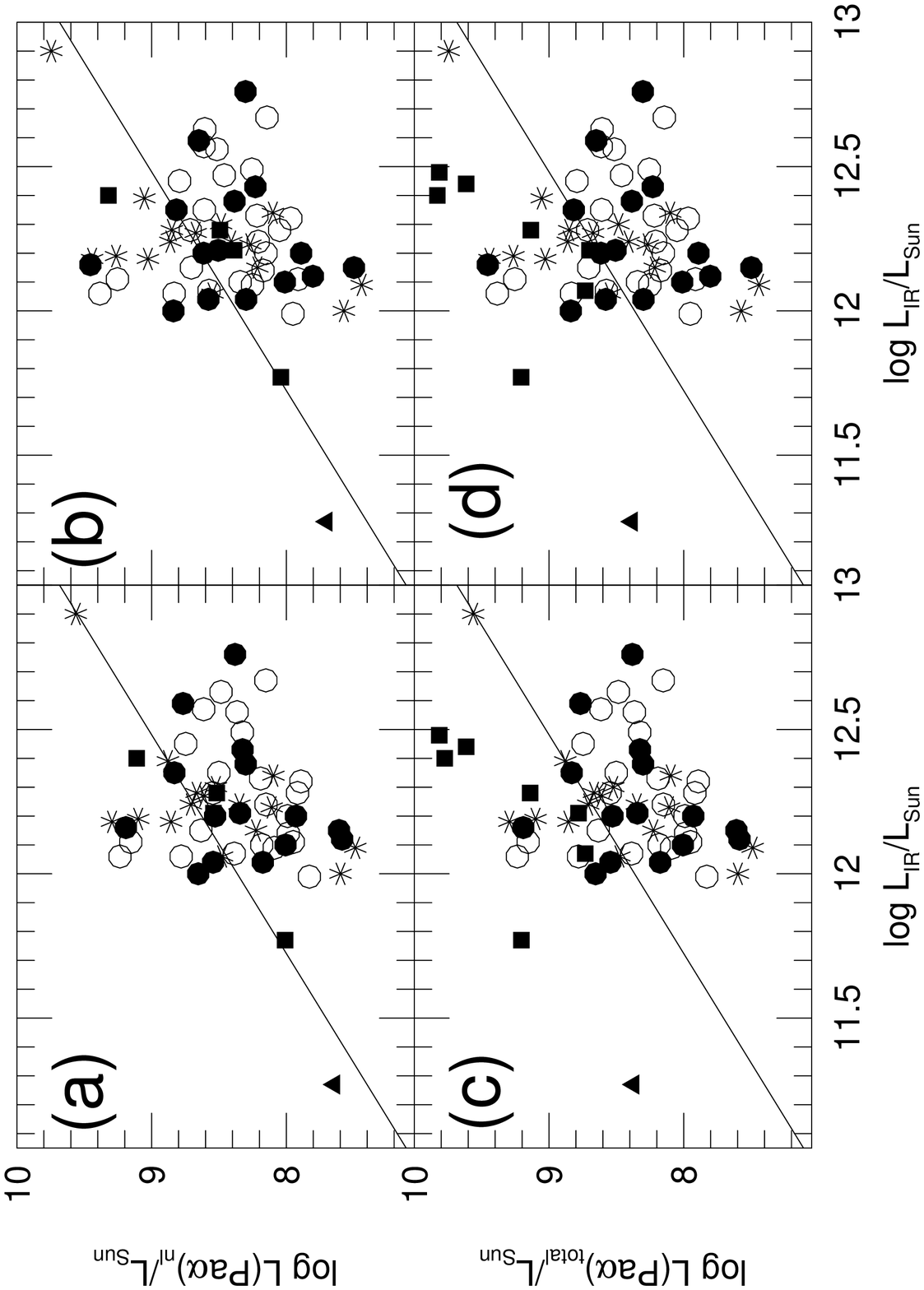}
\caption{ }
\end{figure}

\begin{figure}
%\epsscale{1.2}
\plotone{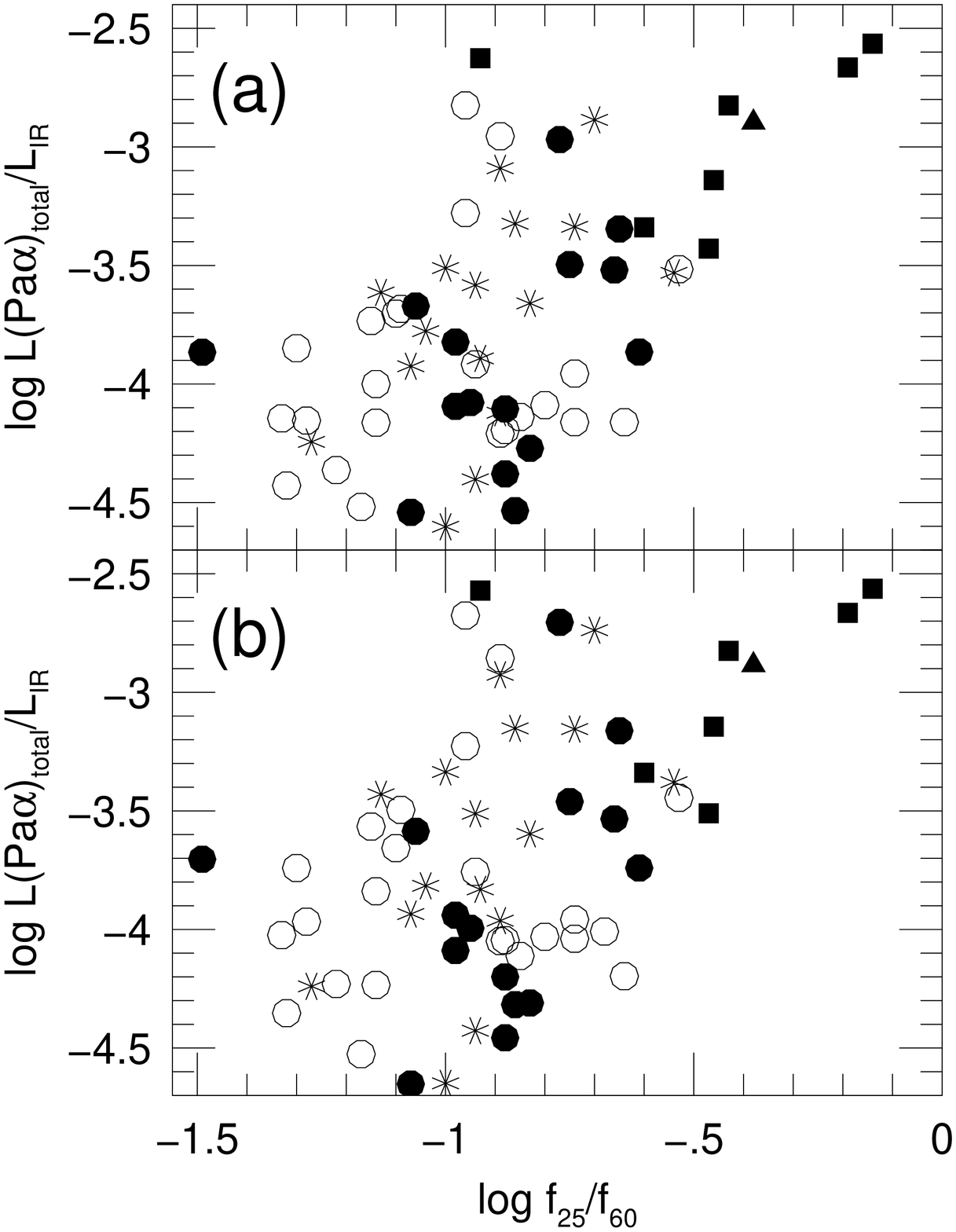}
\caption{ }
\end{figure}

\begin{figure}
%\epsscale{1.2}
\plotone{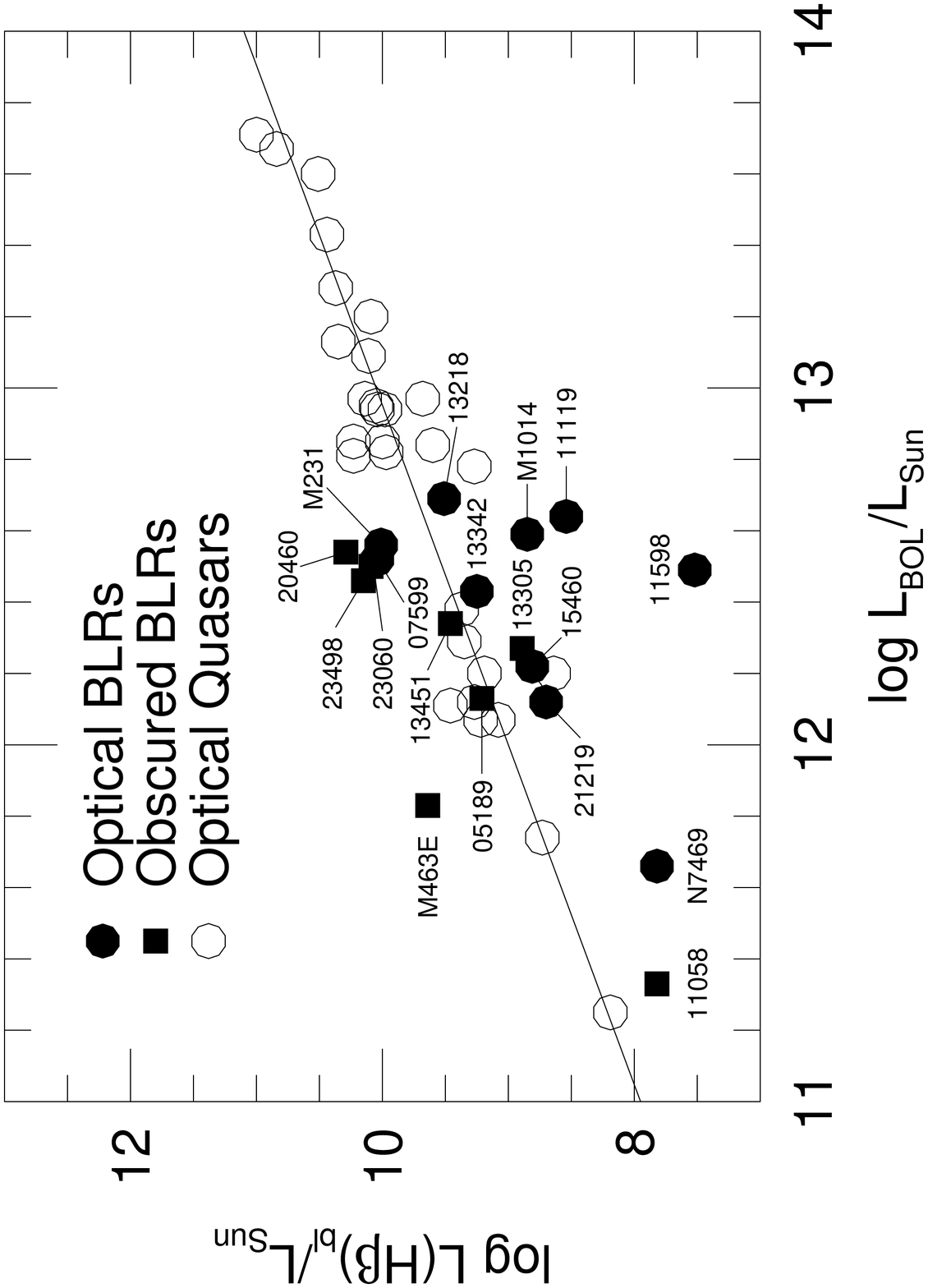}
\caption{ }
\end{figure}


\clearpage

\normalsize

\begin{references}
\refpar
Antonucci, R. 1993, \araa, 31, 473
\refpar
Armus, L., Heckman, T. M., \& Miley, G. K. 1988, \apj, 326, L45
\refpar
Elvis, M. et al. 1994, \apjs, 95, 1
\refpar
Genzel, R., et al. 1998, \apj, 498, 579
\refpar
Goldader, J. D., Joseph, R. D., Doyon, R., \& Sanders, D. B. 1995, \apj, 
444, 97
\refpar
------. 1997a, \apjs, 108, 449
\refpar
------. 1997b, \apj, 474, 104
\refpar
Goodrich, R. W., Veilleux, S., \& Hill, G. J. 1994, \apj, 422, 521
\refpar
Helou, G., Kahn, I. R., Malek, L., \& Boehmer, L. 1988, \apjs, 68, 151
\refpar
Kim, D.-C. 1995, Ph.D. Thesis, University of Hawaii
\refpar
Kim, D.-C., \& Sanders, D. B. 1998, \apjs, 119, 41
\refpar
Kim, D.-C., Sanders, D. B., Veilleux, S., Mazzarella, J. M., \& Soifer, B. T.
1995, \apjs, 98, 129
\refpar
Kim, D.-C., Veilleux, S., \& Sanders, D. B. 1998, \apj, 508, 627
\refpar
Lutz, D., Veilleux, S., \& Genzel, R. 1999, \apj, submitted
\refpar
Marconi, A., Moorwood, A. F. M., Salvati, M., \& Oliva, E. 1994, A\&A, 291, 18
\refpar
Moshir, M., et al. 1992, Explanatory Supplement to the IRAS Faint Source 
Survey, Version 2, JPL~D-10015~8/92 (Pasadena: JPL) (FSC)
\refpar
Mountain, C. M., Robertson, D. J., Lee, T. J., \& Wade, R. 1990, 
Instrumentation in Astronomy VII, ed. D. L. Crawford (Proc. SPIE, Vol. 1235), 
25
\refpar
Mulchaey, J. S., et al. 1994, \apj, 436, 586
\refpar
Osterbrock, D. E., \& de Robertis, M. M. 1985, \pasp, 97, 1129
\refpar
Sanders, D. B., \& Mirabel, I. F. 1996, \araa, 34, 725
\refpar
Sanders, D. B., Phinney, E. S., Neugebauer, G., \& Matthews, K. 1989, \apj, 347, 29
\refpar
Sanders, D. B., Soifer, B. T., Elias, J. H., Madore, B. F., Matthews, K., 
Neugebauer, G., \& Scoville, N. Z. 1988a, \apj, 325, 74
\refpar
Veilleux, S., Goodrich, R. W., \& Hill, G. J. 1997a, \apj, 477, 631
\refpar
Veilleux, S., Kim, D.-C., \& Sanders, D. B. 1999, \apj, submitted (VKS)
\refpar
Veilleux, S., Kim, D.-C., Sanders, D. B., Mazzarella, J. M, \& Soifer, 
B. T. 1995, \apjs, 98, 171
\refpar
Veilleux, S., Sanders, D. B., \& Kim, D.-C. 1997b, \apj, 484, 92
\refpar
Ward, M. J., Blanco, P. R., Wilson, A. S., \& Nishida, M. 1991, \apj, 382, 115
\refpar
Yee, H. K. C. 1980, \apj, 241, 894
\refpar
Young, S., Hough, J. H., Bailey, J. A., Axon, D. J., \& Ward, M. J. 1993, 
\mnras, 260, L1

\end{references}
\end{document}